\documentclass[onecolumn,aps,prd]{revtex4-1}
\RequirePackage[english]{babel}
\RequirePackage[latin1]{inputenc}
\RequirePackage[T1]{fontenc}
\usepackage{graphicx,epsfig}
\usepackage{amsmath}
\usepackage {amssymb}
\usepackage {longtable}
\usepackage{multirow}

\usepackage{subfigmat}
\usepackage{BOONDOX-cal}

\usepackage{dsfont}
\usepackage{dcolumn}
\usepackage{bm}
\usepackage{amsfonts}
\usepackage{subfigure}
\usepackage{color}

\usepackage{BOONDOX-cal}

\def\be{\begin{equation}}
\def\ee{\end{equation}}
\def\beq{\begin{eqnarray}}
\def\eeq{\end{eqnarray}}

\begin{document}
\title{Maximal extension of the Schwarzschild
metric: From Painlev\'e-Gullstrand to Kruskal-Szekeres}




\author{Jos\'{e} P. S. Lemos}
\affiliation{Centro de Astrof\'{\i}sica e Gravita\c c\~ao - CENTRA,
Departamento de F\'{\i}sica, Instituto Superior T\'ecnico - IST,
Universidade de Lisboa - UL, Av. Rovisco Pais 1, 1049-001 Lisboa,
Portugal, email: joselemos@ist.utl.pt}

\author{Diogo L. F. G. Silva}
\affiliation{Centro de Astrof\'{\i}sica e Gravita\c c\~ao - CENTRA,
Departamento de F\'{\i}sica, Instituto Superior T\'ecnico - IST,
Universidade de Lisboa - UL, Av. Rovisco Pais 1, 1049-001 Lisboa,
Portugal, email: diogo.l.silva@tecnico.ulisboa.pt}

\begin{abstract}

We find a specific coordinate system that goes from the
Painlev\'e-Gullstrand partial extension to the Kruskal-Szekeres maximal
extension and thus exhibit the maximal extension of the Schwarzschild
metric in a unified picture. We do this by adopting two time coordinates,
one being the proper time of a congruence of outgoing timelike
geodesics, the other being the proper time of a congruence of ingoing
timelike geodesics, both parameterized by the same energy per unit
mass $E$.  $E$ is in the range $1\leq E<\infty$ with the limit
$E=\infty$ yielding the Kruskal-Szekeres maximal extension. So,
through such an
integrated description one sees that the Kruskal-Szekeres
solution belongs to this family of extensions parameterized by $E$.
Our family of extensions is different from the Novikov-Lema\^itre
family parameterized also by the energy $E$ of timelike geodesics,
with the Novikov extension holding for $0<E<1$ and being maximal, and
the Lema\^itre extension holding for $1\leq E<\infty$ and being
partial, not maximal, and moreover its $E=\infty$ limit evanescing in
a Minkowski spacetime rather than ending in the Kruskal-Szekeres
spacetime.

\end{abstract}


\maketitle



\newpage

\section{Introduction}

The maximal analytical extension of the Schwarzschild solution was a
remarkable achievement in general relativity and in the theory of
black holes.  For the first time the complex causal structure with a
convoluted spacetime topology, stemming from the seemingly trivial
generalization into general relativity of a point particle attractor
in Newtonian gravitation, was unfolded.

It all started with the spherically symmetric vacuum solution of general
relativity found by Schwarzschild \cite{schwarzschild},
that was put in
different terms and in a somewhat different coordinate system by
Droste \cite{droste} and Hilbert \cite{hilbert}, and shown
to be unique by Birkhoff \cite{birkhoff}.  Leaving aside
Schwarzschild's interpretation of Schwarzschild's solution, it is the
solution that later gave rise to black holes. To finalize its full
meaning it was necessary to understand the sphere $r=2M$ that
naturally appears in the solution and accomplish its maximal
extension, i.e., finding the corresponding spacetime in which every
geodesic originating from an arbitrary point in it has infinite length
in both directions or ends at a singularity that cannot be removed by
a coordinate transformation.  These were two problems that proved
difficult.

An early attempt to eliminate the $r=2M$ sphere obstacle and its
inside was provided by Einstein and Rosen \cite{er} that tried to join
smoothly at $r=2M$ two distinct spacetime sheets in order to get some
kind of fundamental particle, in what is known as an Einstein-Rosen
bridge.  Misner and Wheeler \cite{misnerwheeler} generalized the
bridge into a wormhole with a throat at its maximum opening. Wormholes
became a focus of study within general relativity after Morris and
Thorne \cite{thornemorris} showed that with some suitable form of
matter, albeit exotic, they could be traversable, see also, e.g., the
work of Lemos, Lobo and Oliveira \cite{lemoslq}.
The Einstein-Rosen
bridge in terms of the understanding of the $r=2M$ sphere was a dead
end, but as a nontraversable wormhole it reincarnated in the maximal
extensions of the Schwarzschild metric, and as a traversable wormhole
it can be  put in firm ground
once one properly defines it in order to have an admissible
matter support, as disclosed by 
Guendelman, Nissimov,
Pacheva, and Stoilov \cite{guendelman}.

A promising way of seeing the Schwarzschild solution, whatever the
motivation, came with Painlev\'e \cite{painleve} that changed the
Schwarzschild time coordinate into the proper time of a congruence of
ingoing timelike geodesics, or equivalently of ingoing test particles
planted over them, with energy per unit mass $E$ equal to one, that
admitted to put the line element in a new form that was not singular at
$r=2M$. This procedure was also discovered by Gullstrand
\cite{gullstrand}, and the resulting line element, which works as for
outgoing as for ingoing timelike geodesics, is called the
Painlev\'e-Gullstrand line element of the Schwarzschild solution, or
simply referred as Painlev\'e-Gullstrand solution, and in both forms
it is an analytical extension, although partial, of the original
Schwarzschild solution.  The generalization of this line element to
accommodate a congruence of timelike geodesics with any $E$, less or
greater than one, was given by Gautreau and Hoffmann
\cite{gautreauhoffmann}.  The Painlev\'e-Gullstrand line element, not
being singular at $r=2M$, is useful in many understandings of black
hole physics.  For instance, it has been used by Parikh and Wilczek to
understand how Hawking radiation proceeds \cite{parikhwilczek}, or as
a guide for a better understanding of the $r=2M$ sphere by Martel and
Poisson \cite{martelpoisson}, or to understand in new ways the Kerr
metric by Nat\'ario \cite{natario}, or as a generalized slicing of the
Schwarzschild spacetime by Finch \cite{finch} and MacLaurin
\cite{maclaurin2019}.

An extension of the Painlev\'e-Gullstrand line element was given by
Lema\^itre \cite{lemaitre} that transformed the time and radial
coordinates of the Schwarzschild solution to the proper time of
ingoing timelike geodesics with $E=1$ and to a suitable new comoving
radial coordinate, and showed in a stroke that $r=2M$ was a fine
sphere, with nothing singular about it, performing thus an analytical
extension, although partial, of the Schwarzschild solution.  Novikov
\cite{novikov} understood that for timelike geodesics with $0<E<1$ it
was possible to perform a maximal analytical extension and display the
Schwarzschild solution in its fullness. The Lema\^itre extension, as
an exterior spacetime, was implicitly used in the gravitational
contraction of a cloud of dust by Oppenheimer and Snyder to discover
black holes and their formation for the first time with the natural
appearance of an exterior
event horizon at $r=2M$ \cite{oppsny}.  Presentations
of the Novikov-Lemaitre extensions can be seen in several places.  The
Novikov maximal extension is worked through in Zel'dovich and
Novikov's book \cite{zeldovichnovikov} and in Gautreau
\cite{gautreau1980}, and the Lema\^itre extension is featured, e.g.,
in the detailed book by Krasi\'nski \cite{krasinski} and in the very
useful book of Blau \cite{blau2014}.

Remarkably, there is a parallel development that uses lightlike, or
null, geodesics rather than timelike ones.  Indeed, Eddington
\cite{eddington} used ingoing null geodesics to transform the
Schwarzschild time into a new time that straightened out those very
ingoing null geodesics and to put the line element in a new form that
was not singular at $r=2M$.  This was recovered by Finkelstein
\cite{finkelstein}, and then Penrose \cite{penrose} understood that it
was more natural to use the corresponding advanced null coordinate to
represent the metric and the line element. This form works as for
outgoing as for ingoing null geodesics, and the solution is
correspondingly called the Schwarzschild solution in retarded or in
advanced null Eddington-Finkelstein coordinates, respectively. Both
forms are analytical extensions, although partial, of the original
Schwarzschild solution.  The Eddington-Finkelstein line element, not
being singular at $r=2M$ is also useful in many understandings of
black hole physics.  For instance, it has been used by Alcubierre and
Bruegmann in black hole excision in 3+1 numerical relativity
\cite{alcubierre}, or as a guide for a better understanding of the
$r=2M$ sphere by Adler, Bjorken, Chen, and Liu \cite{adler}, or to
understand perturbatively the accretion of matter onto a black hole
\cite{bde}, or to understand the 
stress-energy tensor of quantum fields involved in the
evaporation of a black hole \cite{abpagetzou}, or even to
treat quantum gravitational problems
related to coordinate transformations \cite{goodunruh}.

An extension to the Eddington-Finkelstein line element was given by
Kruskal \cite{kruskal} and Szekeres \cite{szekeres}. By using both
outgoing and ingoing null geodesics to transform the Schwarzschild
time and the Schwarzschild radius
coordinates into new analytical extended
time and spatial coordinates, both the outgoing and the ingoing null
geodesics were straightened out and in addition one could pass with
ease the sphere $r=2M$ in all directions. In this way the maximal
analytical extension of the Schwarzschild solution was unfolded, in a
single coordinate system, into its full form.  Fuller and Wheeler
\cite{fullerwheeler} revealed its dynamic structure with a
nontraversable Einstein-Rosen bridge, i.e., a nontraversable wormhole,
lurking in-between two distinct asymptotically flat spacetime regions
and driving, out of spacetime spacelike singularities at $r=0$, the
creation of a white hole into the formation of a black hole.
Prior maximal extensions had also been given in Synge
\cite{synge} and Fronsdal \cite{fronsdal} using several coordinate
systems or embeddings, rather than the unique coordinate system of the
Kruskal-Szekeres extension.
Modern
presentations of the Kruskal-Szekeres solution can be seen in the
books on general relativity and gravitation
by Hawking and Ellis \cite{hawking}, Misner, Thorne, and Wheeler
\cite{mtw}, Wald \cite{wald}, d'Inverno \cite{dinverno},
Bronnikov and Rubin
\cite{bronn}, and
Chru\'sciel \cite{crusciel}, and in
many other places, where double null coordinates are usually
employed.
The Kruskal-Szekeres line element, with its maximal properties, 
is certainly useful in a great very many understandings of
black hole physics, notably, it surely is a
prototype of gravitational collapse. 
To name two further examples of its applicability, 
Zaslavskii \cite{zaslavskii}
has used its properties
to suitably define high energy collisions
in the vicinity of the event horizon,
and 
Hodgkinson, Louko, and
Ottewill \cite{louko}
have examined the response of particle detectors
to fields 
in diverse quantum vacuum states
working with Kruskal-Szekeres spacetime and coordinates.

Now, the Painlev\'e-Gullstrand line element uses as coordinate the
proper time of a congruence of outgoing or ingoing timelike geodesics
and the Eddington-Finkelstein line element uses as coordinate the
retarded or advanced null parameter of a congruence of outgoing or ingoing
null geodesics, respectively.
There is a connection between the
two coordinate systems as worked out by Lemos \cite{jpsl},
who showed that by taking the $E=\infty$ limit of the
Painlev\'e-Gullstrand line element, and more generally its
Lema\^itre-Tolman-Bondi generalization to include dust matter, one
obtains the Eddington-Finkelstein line element, and more generally its
Vaidya generalization to include incoherent radiation.  Indeed, since
$E$ is the energy per unit mass of the timelike geodesic,
or of the particle placed over it,
when the
mass goes to zero, $E$ goes to infinity, and the proper time along the
timelike geodesic turns into a well defined affine parameter along the
null geodesic, or along the lightlike particle trajectory
placed over it.

But now we have a conundrum.  The Novikov-Lemaitre family of solutions
parameterized by $E$ comes out of the corresponding
Painlev\'e-Gullstrand family with the addition of an appropriate
radial coordinate. On the one hand, the Novikov solution is maximal,
on the other hand, the Lema\^itre solution is not. Moreover, although
Painlev\'e-Gullstrand goes into Eddington-Finkelstein in the
$E=\infty$ limit, Lema\^itre does not go into
Kruskal-Szekeres in the $E=\infty$ limit, instead it dies in a
Minkowski spacetime.  But Eddington-Finkelstein goes
into Kruskal-Szekeres.  In brief, Painlev\'e-Gullstrand goes into
Novikov-Lema\^itre that does not go into Kruskal-Szekeres, and
Painlev\'e-Gullstrand goes into Eddington-Finkelstein that goes into
Kruskal-Szekeres.  So, there is a missing link. What is the maximal
extension that starts from Painlev\'e-Gullstrand and in the $E=\infty$
limit goes into the Kruskal-Szekeres maximal extension?

Here, we find the maximal analytic extension of the Schwarzschild
spacetime that goes from Painlev\'e-Gullstrand to Kruskal-Szekeres
yielding a unified picture
of extensions.  By using two analytically
extended Painlev\'e-Gullstrand time coordinates, we find
another way of obtaining the maximal analytic extension of the
Schwarzschild spacetime.  It is parameterized by the energy $E$ of the
outgoing and ingoing timelike geodesics.  The extension is valid for
$1\leq E<\infty$, with the case $E=\infty$ giving the
Kruskal-Szekeres extension.  So the Kruskal-Szekeres extension is a
member of this family.  It is a different family from the
Novikov-Lema\^itre family, which does not have as its member the
Kruskal-Szekeres extension, and moreover the $E\geq1$ Lema\^itre
extension is not maximal.  It is certainly
opportune to incorporate into a family of
maximal $E$ extensions of the Schwarzschild metric, the maximal extension
of Kruskal and Szekeres in the year we celebrate its 60 years.

The paper is organized as follows. In Sec.~\ref{doublePG},
we give the Schwarzschild metric in double  Painlev\'e-Gullstrand
coordinates for $E>1$.
In Sec.~\ref{E>1m}, we extend the Schwarzschild metric for $E>1$
past the $r=2M$ coordinate singularity using analytical extended 
coordinates, and produce its maximal analytical
extension. 
In Sec.~\ref{E=1m}, we give the $E=1$  maximal analytical
extension as the limit from $E>1$.
In Sec.~\ref{E=oom}, we give the $E=\infty$  maximal analytical
extension as the limit from $E>1$ and show that it is
the Kruskal-Szekeres maximal extension.
In Sec.~\ref{allEm}, we present the causal structure of
the maximal extended spacetime for several $E$, from
$E=1$ to $E=\infty$.
In Sec.~\ref{conc}, we conclude.
In the Appendix, we show in detail
the limits $E=1$ and
$E=\infty$ directly
from the $E>1$ generic case.

\vfill

\section{The Schwarzschild solution in double Painlev\'e-Gullstrand
form}
\label{doublePG}

The vacuum Einstein equation $G_{ab}=0$, where
$G_{ab}$ is the Einstein tensor and $a,b$ are spacetime
indices, give for a line element
$ds^2=g_{ab}(x^a)\hskip0.01cm dx^adx^b$,
where $g_{ab}(x^a)$ is the metric and $x^a$ are the coordinates,
in the classical standard spherical symmetric
coordinates $(t,r,\theta,\phi)$
the Schwarzschild solution, namely,
\begin{equation}
\label{eqn:Schmetric}
\mathrm{d}s^2=-\left( 1-\frac{2M}{r}\right)\mathrm{d}t^2+
\dfrac{dr^2}{1-\frac{2M}{r}} +r^2
(\mathrm{d} \theta^2+ \sin^2{\theta}\,\mathrm{d} \phi^2)\,,
\end{equation}
where $M$ is the spacetime mass.
We assume $M\geq0$ and $r\geq0$.
In this form the line element, and so the metric, is
singular at the, Schwarzschild, gravitational,
or event horizon radius $r=2M$, and at $r=0$.
For $r>2M$, the Schwarzschild coordinate $t$
is timelike and the coordinate $r$ is spacelike,  a 
radial coordinate.
For $r<2M$, these coordinates swap roles,
the Schwarzschild coordinate $t$
is spacelike
and the coordinate $r$ is timelike.

We now apply a first coordinate transformation
such that the Schwarzschild time $t$
in Eq.~(\ref{eqn:Schmetric}) goes into a new time
${\mathcal t}={\mathcal t}(t,r)$ given in differential
form by
\begin{equation}
\label{eqn:coordtransf2}
\mathrm{d}{\mathcal t}=
E\mathrm{d}t-
\dfrac{\left( E^2-1+ \frac{2M}{r} \right)^{1/2}}
{1- \frac{2M}{r}}
\mathrm{d}r\,,
\end{equation}
with $E\geq1$, $E$ being a parameter.
This is a Painlev\'e-Gullstrand
coordinate transformation for the
congruence of outgoing radial timelike
geodesics with energy $E$. 
We can also perform a different coordinate transformation,
such that the Schwarzschild time $t$ in Eq.~(\ref{eqn:Schmetric})
goes into a new time
$\tau=\tau(t,r)$ given in differential
form by
\begin{equation}
\label{eqn:coordtransf1}
\mathrm{d}\tau={E} \mathrm{d}t+
\dfrac{\left( E^2-1+ \frac{2M}{r} \right)^{1/2}}
{1- \frac{2M}{r}}
\mathrm{d}r\,.
\end{equation}
with $E\geq1$, $E$ being the same parameter as above.
This is a Painlev\'e-Gullstrand
coordinate transformation for the
congruence of ingoing radial timelike
geodesics with energy $E$.
The two transformations together, ${\mathcal t}={\mathcal t}(t,r)$ 
and $\tau=\tau(t,r)$,
Eqs.~(\ref{eqn:coordtransf2}) and (\ref{eqn:coordtransf1}),
respectively,
can then be seen
as a transformation from the  Schwarzschild time
and radius $(t,r)$ to the two new coordinates
$({\mathcal t},\tau)$.
The inverse transformations,
from $({\mathcal t},\tau)$ to $(t,r)$, in differential
form are 
\begin{equation}
\label{inverse1}
E\mathrm{d}t= \frac12\left(\;
\mathrm{d}{\mathcal t}+\mathrm{d}\tau
\right)\,,
\end{equation}
\begin{equation}
\label{inverse2}
\frac{\left( E^2-1+ \frac{2M}{r} \right)^{1/2}}
{\left( 1- \frac{2M}{r} \right)}
\mathrm{d}r= \frac12\left(-\mathrm{d}{\mathcal t}+\mathrm{d}\tau
\right)\,.
\end{equation}

Applying the coordinate transformation given in
Eq.~(\ref{eqn:coordtransf2}) to the Schwarzschild line element,
Eq.~(\ref{eqn:Schmetric}), gives the  line element in Painlev\'e-Gullstrand
outgoing coordinates with energy parameter $E\geq1$, namely,
$\mathrm{d}s^2=- \frac{1}{E^2} \left( 1- \frac{2M}{r} \right)
\mathrm{d}{\mathcal t}^2- \allowbreak - 2 \frac{1}{E^2} \sqrt{E^2-1+ \frac{2M}{r}}
\mathrm{d}{\mathcal t}\, \mathrm{d}r+ \frac{1}{E^2} \mathrm{d}r^2+
r^2( \mathrm{d}\theta^2+\sin^2{\theta}\, \mathrm{d}\phi^2)$.  This
form of the metric is not singular anymore at $r=2M$, but there is
still the singularity at $r=0$ which cannot be removed. Note that
inside $r=2M$ this Painlev\'e-Gullstrand form has the feature of
having two time coordinates, ${\mathcal t}$ and $r$.
Applying the coordinate transformation given in
Eq.~(\ref{eqn:coordtransf1}) to the Schwarzschild metric,
Eq.~(\ref{eqn:Schmetric}), gives the metric in Painlev\'e-Gullstrand
ingoing coordinates with energy parameter $E\geq1$, namely,
$\mathrm{d}s^2=- \frac{1}{E^2} \left( 1- \frac{2M}{r} \right)
\mathrm{d}\tau^2+ 2 \frac{1}{E^2} \sqrt{E^2-1+ \frac{2M}{r}}
\mathrm{d}\tau\, \mathrm{d}r+ \frac{1}{E^2} \mathrm{d}r^2+ r^2(
\mathrm{d}\theta^2+\sin^2{\theta}\, \mathrm{d}\phi^2)$. This form of
the metric is also not singular anymore at $r=2M$, but there is still the
singularity at $r=0$ which cannot be removed. Note that inside $r=2M$
this Painlev\'e-Gullstrand form has the feature of having two time
coordinates, $\tau$ and $r$. All of this is well known.

We now apply a simultaneous coordinate transformation, given through
Eqs.~(\ref{eqn:coordtransf2})-(\ref{eqn:coordtransf1}),
or if
one prefers
Eqs.~(\ref{inverse1})-(\ref{inverse2}), to the Schwarzschild metric,
Eq.~(\ref{eqn:Schmetric}), to get
\begin{equation}
\label{eqn:genmetric0}
\begin{split}
\mathrm{d}s^2&= -\frac{1}{4E^2}
\frac{1-\frac{2M}{r}}{E^2-1+\frac{2M}{r}}
\left[ -\left( 1- \frac{2M}{r} \right)
(\mathrm{d}{\mathcal t}^2+\mathrm{d}\tau^2)+
2 \left( 2E^2-1+ \frac{2M}{r}
\right) \mathrm{d}{\mathcal t}\,\mathrm{d}\tau  \right]+\\ &+
r^2({\mathcal t},\tau) (\mathrm{d}\theta^2+
\sin^2{\theta}\, \mathrm{d}\phi^2)\,,
\end{split}
\end{equation}
with $r({\mathcal t},\tau)$ obtained via
Eq.~(\ref{inverse2})
and depends on whether $E=1$ or $E>1$.
This is the Schwarzschild metric in double
Painlev\'e-Gullstrand coordinates.

The line element of Eq.~(\ref{eqn:genmetric0}) is still degenerate for
$r=2M$. So, if we want to extend it past this sphere
we have to perform another set of coordinate
transformations.  This set is given by
$\frac{{\mathcal t}'}{M}= -\exp{\left( -\frac{{\mathcal t}}{4ME}
\right) }$
and $\frac{\tau'}{M}=  \exp{\left( \;\frac{\tau}{4ME} \right) }$.
When applied to 
Eq.~(\ref{eqn:genmetric}), it gives,
$\mathrm{d}s^2= 4M^2 \frac{1-\frac{2M}{r}}{E^2-1+\frac{2M}{r}}
\left[
\left(1-\frac{2M}{r} \right)
\left(\frac{\mathrm{d}{\mathcal t}'^2}{{\mathcal t}'^2}+
\frac{\mathrm{d}\tau'^2}{\tau'^2} \right)+
2\left(2E^2-1+\frac{2M}{r}\right)
\frac{\mathrm{d}{\mathcal t}'}{{\mathcal t}'}\,
\frac{\mathrm{d}\tau'}{\tau'}
\right]
+ \allowbreak + r^2({\mathcal t'},\tau')(\mathrm{d}\theta^2+ \sin^2{\theta}\,
\mathrm{d}\phi^2)$, with $r({\mathcal t'},\tau')$
a function that is given implicitly.
The form of this metric will depend on the value of $E$
through the solution to the differential coordinate relations,
Eqs.~(\ref{eqn:coordtransf2}) and (\ref{eqn:coordtransf1}),
or equivalently, Eqs.~(\ref{inverse1})-(\ref{inverse2}). Clearly,
the case $E<1$ cannot be treated from the formulas above
and we have dismissed it from the start.
Therefore we restrict the analysis to $1\leq E<\infty$.
The $E=1$ and $E=\infty$ can be seen as
limiting cases of the 
generic $E>1$ case. 
Let us do the $E>1$ case in detail and then treat
$E=1$ and $E=\infty$ as the inferior and superior
limiting cases, respectively, 
of $E>1$.



\section{Maximal analytic extension for $E>1$ as generic case}
\label{E>1m}

To start building the maximal analytic extension for $E>1$,
we find the solutions to
the new coordinates ${\mathcal t}$ and $\tau$
from Eqs.~(\ref{eqn:coordtransf2}) and
(\ref{eqn:coordtransf1}).
When $E>1$ they are
\begin{align}
\label{eqn:closedrel2gt1}
\begin{split}
{\mathcal t}&=Et- r\sqrt{E^2-1+ \frac{2M}{r}}
-
2ME \ln{\left|
\frac{2M}{r}\left(
\frac{\frac{r}{2M}-1}
{2E^2-1+\frac{2M}{r}+2E\sqrt{E^2-1+\frac{2M}{r}}}
\right)\right|}-
\\ &
-
M\frac{2E^2-1}{\sqrt{E^2-1}}
\ln\left[
\frac{r}{M}
\left(
\sqrt{E^2-1}
\sqrt{E^2-1+\frac{2M}{r}}+E^2-1+\frac{M}{r}
\right)\right]
\,,
\end{split}
\\
\label{eqn:closedrel1gt1}
\begin{split}
\tau&=Et+ r\sqrt{E^2-1+ \frac{2M}{r}}
+
2ME \ln{\left|
\frac{2M}{r}\left(
\frac{\frac{r}{2M}-1}
{2E^2-1+\frac{2M}{r}+2E\sqrt{E^2-1+\frac{2M}{r}}}
\right)\right|}+
\\ &
+
M\frac{2E^2-1}{\sqrt{E^2-1}}
\ln\left[
\frac{r}{M}
\left(
\sqrt{E^2-1}
\sqrt{E^2-1+\frac{2M}{r}}+E^2-1+\frac{M}{r}
\right)\right]
\,.
\end{split}
\end{align}
The line element to start with is 
\begin{equation}
\label{eqn:genmetric}
\begin{split}
\mathrm{d}s^2&= -\frac{1}{4E^2} \frac{1-
\frac{2M}{r}}{E^2-1+\frac{2M}{r}}
\left[ -\left( 1- \frac{2M}{r} \right)
(\mathrm{d}{\mathcal t}^2+\mathrm{d}\tau^2)+
2 \left( 2E^2-1+ \frac{2M}{r}
\right) \mathrm{d}{\mathcal t}\,\mathrm{d}\tau  \right]+\\ &+
r^2 (\mathrm{d}\theta^2+ \sin^2{\theta}\, \mathrm{d}\phi^2)\,,
\end{split}
\end{equation}
which is taken from
Eq.~(\ref{eqn:genmetric0}),
now bearing in mind that $E>1$ implicitly here, 
and
with $r=r({\mathcal t},\tau)$ being
obtained via Eqs.~(\ref{eqn:closedrel2gt1})
and (\ref{eqn:closedrel1gt1}), i.e.,
\begin{eqnarray}
\label{rimplicitE>1}
r\sqrt{E^2-1+ \frac{2M}{r}}
+2ME \ln{\left|
\frac{2M}{r}\left(
\frac{\frac{r}{2M}-1}
{2E^2-1+\frac{2M}{r}+2E\sqrt{E^2-1+\frac{2M}{r}}}
\right)\right|}+\nonumber
\\
M\frac{2E^2-1}{\sqrt{E^2-1}}
\ln\left[
\frac{r}{M}
\left(
\sqrt{E^2-1}
\sqrt{E^2-1+\frac{2M}{r}}+E^2-1+\frac{M}{r}
\right)\right]
=&\frac12\left(-{\mathcal t}+\tau\right)\,.
\end{eqnarray}

The line element Eq.~(\ref{eqn:genmetric}) is still degenerate at
$r=2M$. So, if we want to extend past it we have to do something.  To
remove this behavior, we proceed with two new coordinate
transformations given by $\frac{{\mathcal t}'}{M}= -\exp{\left(
-\frac{{\mathcal t}}{4ME}\right)}$ and $\frac{\tau'}{M}= \quad
\exp{\left( \;\frac{\tau}{4ME} \right)}$, for $r>2M$.  Then, using
Eqs.~(\ref{eqn:closedrel2gt1}) and (\ref{eqn:closedrel1gt1}) the
maximal extended coordinates 
${\mathcal t}'$  and
$\tau'$ are
\begin{align}
\label{eqn:coordtransfrho2gt1}
\begin{split}
\frac{{\mathcal t}'}{M}&= -\exp{\left( -\frac{{\mathcal t}}{4ME}
\right) }\,, \quad {\rm i.e.,} \quad\\
\frac{{\mathcal t}'}{M}&=
-\sqrt{\frac{2M}{r}}
\frac{\sqrt{\frac{r}{2M}-1}}
{\sqrt{2E^2- 1+\frac{2M}{r}+ 2E\sqrt{E^2-1+\frac{2M}{r}}}
}
\exp{\left( -\frac{t}{4M}
+\frac{r}{4ME} \sqrt{E^2-1+ \frac{2M}{r}}\right)}
\times
\\ &
\times
\left[
\frac{r}{M}
\left(
\sqrt{E^2-1} \sqrt{E^2-1+\frac{2M}{r}}+ E^2-1+\frac{M}{r}
\right)\right]^{\frac{2E^2-1}{4E \sqrt{E^2-1}}}
\,,
\end{split}\\
\label{eqn:coordtransftau2gt1}
\begin{split}
\frac{{\tau}'}{M}&= \exp{\left( \frac{\tau}{4ME}
\right) }\,, \quad {\rm i.e.,} \quad\\
\frac{\tau'}{M}&=
\sqrt{\frac{2M}{r}}
\frac{\sqrt{\frac{r}{2M}-1}}
{\sqrt{2E^2- 1+\frac{2M}{r}+ 2E\sqrt{E^2-1+\frac{2M}{r}}}
}
\exp{\left( \frac{t}{4M} +\frac{r}{4ME} \sqrt{E^2-1+
\frac{2M}{r}}\right)}
\times
\\ &
\times
\left[
\frac{r}{M}
\left(
\sqrt{E^2-1} \sqrt{E^2-1+\frac{2M}{r}}+ E^2-1+\frac{M}{r}
\right)\right]^{\frac{2E^2-1}{4E \sqrt{E^2-1}}}
\,,
\end{split}
\end{align}
respectively.
Putting ${\mathcal t}'$  and $\tau'$ given in
Eqs.~(\ref{eqn:coordtransfrho2gt1})
and (\ref{eqn:coordtransftau2gt1}),
respectively, into the line element
Eq.~(\ref{eqn:genmetric}), 
one finds the new line element
in coordinates $({\mathcal t}',\tau',\theta,\phi)$
given by
\begin{equation}
\label{eqn:metricE>1}
\begin{split}
\mathrm{d}s^2&=-
4 \left( \frac{ 2E^2-1+ \frac{2M}{r}+ 2E
\sqrt{E^2-1+\frac{2M}{r}}}{ E^2-1+\frac{2M}{r}} \right)
\exp{\left( -\frac{r}{2ME} \sqrt{E^2-1+ \frac{2M}{r}}
\right)}
\times
\\
&\times \left(\frac{M}{r}\, \frac{1}{E^2-1+ \frac{M}{r}+
\sqrt{E^2-1}\sqrt{E^2-1+\frac{2M}{r}}} \right)
^{\frac{2E^2-1}{2E \sqrt{E^2-1}}}
\times
\\
&\times
\Bigg[ -\frac{1}{M^2} \left(2E^2-1+ \frac{2M}{r}+2E
\sqrt{E^2-1+ \frac{2M}{r}} \right)
\exp{\left( -\frac{r}{2ME} \sqrt{E^2-1+ \frac{2M}{r}}
\right)}
\times
\\
&\times \left(\frac{M}{r}
\frac{1}{E^2-1+\frac{M}{r}+
\sqrt{E^2-1} \sqrt{E^2-1+\frac{2M}{r}}}
\right)^{\frac{2E^2-1}{2E \sqrt{E^2-1}}}
(\tau'^2\mathrm{d}{\mathcal t}'^2+{\mathcal t}'^2
\mathrm{d}\tau'^2)+
\\
&+2 \left( 2E^2-1+ \frac{2M}{r} \right)
\mathrm{d}{\mathcal t}'\mathrm{d}\tau'\,
\Bigg]+ r^2 (\mathrm{d}\theta^2+
\sin^2{\theta}\, \mathrm{d}\phi^2)\,,
\end{split}
\end{equation}
where $r=r({\mathcal t}',\tau')$ is defined implicitly as a function
of ${\mathcal t}'$ and $\tau'$ through
\begin{equation}
\label{rimplicitEoo}
\begin{split}
& 
\left( \frac{
\frac{r}{2M}-1}{2E^2-1+ \frac{2M}{r}+ 2E \sqrt{E^2-1+
\frac{2M}{r}}} \right) \frac{2M}{r}\,
\exp{\left( \frac{r}{2ME}
\sqrt{E^2-1+ \frac{2M}{r}} \right)}
\times \\
& \times
\left[
\frac{r}{M}
\left( E^2-1 + \frac{M}{r}+ \sqrt{E^2-1}
\sqrt{E^2-1+ \frac{2M}{r}}\right)
\right]^{\frac{2E^2-1}{2E \sqrt{E^2-1}}}=
-\frac{{\mathcal t}'}{M}
\frac{\tau'}{M}\,.
\end{split}
\end{equation}
All of this is done so that ${\mathcal t}'$ and $\tau'$ have ranges
$-\infty<{\mathcal t}'<\infty$ and $-\infty<\tau'<\infty$, which
Eqs.~(\ref{eqn:metricE>1}) and (\ref{rimplicitEoo}) permit.  Several
properties are now worth mentioning.

In terms of the coordinates $({\mathcal t},\tau)$, or
$(t,r)$, the coordinate
transformations that yield
the maximal extended coordinates $({\mathcal t}',\tau')$
with infinite ranges have to be broadened, resulting
in the existence of four regions, regions I, II, III, and IV.
Region I is the region where the transformations
Eqs.~(\ref{eqn:coordtransfrho2gt1}) and~(\ref{eqn:coordtransftau2gt1})
hold, i.e., it is a region with ${\mathcal t}'\leq 0$ and $\tau'\geq
0$. It is a region with $r\geq 2M$ and $-\infty<t<\infty$.
Of course, in this region Eqs.~(\ref{eqn:metricE>1}) and (\ref{rimplicitEoo})
hold.
Region II, a region for which $r\leq 2M$, gets a different set of
coordinate transformations.  In this $r\leq 2M$ region, due to the
moduli appearing in Eqs.~(\ref{eqn:closedrel2gt1})
and~(\ref{eqn:closedrel1gt1}) and the change of sign in
Eq.~(\ref{eqn:genmetric}), one defines instead
${\mathcal t}'$ as
$ \frac{{\mathcal t}'}{M}=+ \exp{\left(-\frac{{\mathcal
t}}{4ME}\right)}=\allowbreak
=\sqrt{\frac{2M}{r}} \frac{\sqrt{1-\frac{r}{2M}}}
{\sqrt{2E^2- 1+\frac{2M}{r}+ 2E\sqrt{E^2-1+\frac{2M}{r}}} }
\exp{\left( -\frac{t}{4M} +\frac{r}{4ME} \sqrt{E^2-1+
\frac{2M}{r}}\right)} \left[ \frac{r}{M} \left( \sqrt{E^2-1}
\sqrt{E^2-1+\frac{2M}{r}}+ E^2-1+\right.\right.  \\
\left.\left.+\frac{M}{r} \right)\right]^{\frac{2E^2-1}{4E
\sqrt{E^2-1}}}$
and $\tau'$ as
$\frac{\tau'}{M} = \exp{\left( \frac{\tau}{4ME} \right) }=
\sqrt{\frac{2M}{r}} \frac{\sqrt{1-\frac{r}{2M}}} {\sqrt{2E^2-
1+\frac{2M}{r}+ 2E\sqrt{E^2-1+\frac{2M}{r}}} } \exp{\left(
\frac{t}{4M} +\frac{r}{4ME} \sqrt{E^2-1+ \frac{2M}{r}}\right)} \times \allowbreak
\times \left[ \frac{r}{M} \left( \sqrt{E^2-1} \sqrt{E^2-1+\frac{2M}{r}}+
E^2-1+\frac{M}{r} \right)\right]^{\frac{2E^2-1}{4E \sqrt{E^2-1}}}$.
These transformations are valid for ${\mathcal t}'\geq 0$ and
$\tau'\geq 0$.
It is a region with $r\leq2M$ and $-\infty<t<\infty$.
Note that the coordinate transformations in this region give
$\left( \frac{1- \frac{r}{2M}}{2E^2-1+ \frac{2M}{r}+ 2E \sqrt{E^2-1+
\frac{2M}{r}}} \right) \frac{2M}{r}\, \exp{\left( \frac{r}{2ME}
\sqrt{E^2-1+ \frac{2M}{r}} \right)} \left[ \frac{r}{M} \left( E^2-1 +
\frac{M}{r}+ \sqrt{E^2-1} \sqrt{E^2-1+ \frac{2M}{r}}\right)
\right]^{\frac{2E^2-1}{2E \sqrt{E^2-1}}}=\\
=\frac{{\mathcal t}'}{M}
\frac{\tau'}{M}$.
But all
this has been automatically incorporated
into Eqs.~(\ref{eqn:metricE>1}) and (\ref{rimplicitEoo})
so there is no further concern on that.
Region III is another $r\geq 2M$ region.
Now one defines ${\mathcal t}'$
as
$\frac{{\mathcal t}'}{M}= \exp{\left(-\frac{{\mathcal
t}}{4ME}\right)}= \sqrt{\frac{2M}{r}} \frac{\sqrt{\frac{r}{2M}-1}}
{\sqrt{2E^2- 1+\frac{2M}{r}+ 2E\sqrt{E^2-1+\frac{2M}{r}}}}
\times \\
\times \exp{\left( -\frac{t}{4M} +\frac{r}{4ME} \sqrt{E^2-1+
\frac{2M}{r}}\right)} \left[ \frac{r}{M} \left( \sqrt{E^2-1}
\sqrt{E^2-1+\frac{2M}{r}}+ E^2-1+\frac{M}{r}
\right)\right]^{\frac{2E^2-1}{4E \sqrt{E^2-1}}}$
and $\tau'$ as 
$\frac{\tau'}{M} =\\
=-\exp{\left( \frac{\tau}{4ME} \right) } =-
\sqrt{\frac{2M}{r}} \frac{\sqrt{\frac{r}{2M}-1}} {\sqrt{2E^2-
1+\frac{2M}{r}+ 2E\sqrt{E^2-1+\frac{2M}{r}}} } \exp{\left(
\frac{t}{4M} +\frac{r}{4ME} \sqrt{E^2-1+ \frac{2M}{r}}\right)} \left[
\frac{r}{M} \left( \sqrt{E^2-1} \sqrt{E^2-1+\frac{2M}{r}}+\right.\right.
\\
\left.\left. +
E^2-1 +\frac{M}{r} \right)\right]^{\frac{2E^2-1}{4E
\sqrt{E^2-1}}}$.
These transformations are valid for the region with ${\mathcal t}'\geq
0$ and $\tau'\leq 0$. It is a region with $r\geq\\ \geq 2M$ and
$-\infty<t<\infty$.  Note that the coordinate transformations in this
region give
$\left( \frac{ \frac{r}{2M}-1}{2E^2-1+ \frac{2M}{r}+ 2E \sqrt{E^2-1+
\frac{2M}{r}}} \right) \frac{2M}{r}\times
\\
\times \exp{\left( \frac{r}{2ME} \sqrt{E^2-1+ \frac{2M}{r}} \right)}
\left[ \frac{r}{M} \left( E^2-1 + \frac{M}{r}+ \sqrt{E^2-1}
\sqrt{E^2-1+ \frac{2M}{r}}\right) \right]^{\frac{2E^2-1}{2E
\sqrt{E^2-1}}}=- \frac{{\mathcal t}'}{M} \frac{\tau'}{M}$.
But all
this has been automatically incorporated
into Eqs.~(\ref{eqn:metricE>1}) and (\ref{rimplicitEoo})
so again there is no further concern on that.
\noindent
Region IV is another region with $r\leq 2M$.
Now, one defines ${\mathcal t}'$
as
$\frac{{\mathcal t}'}{M}=- \exp{\left(-\frac{{\mathcal
t}}{4ME}\right)}=- \sqrt{\frac{2M}{r}} \frac{\sqrt{1-\frac{r}{2M}}}
{\sqrt{2E^2- 1+\frac{2M}{r}+ 2E\sqrt{E^2-1+\frac{2M}{r}}} }
\exp{\left( -\frac{t}{4M}+ \right.}\allowbreak
{\left.+\frac{r}{4ME} \sqrt{E^2-1+
\frac{2M}{r}}\right)}
\left[
\frac{r}{M}
\left( \sqrt{E^2-1}
\sqrt{E^2-1+\frac{2M}{r}}+
E^2-1+\frac{M}{r} \right)\right]^{\frac{2E^2-1}{4E
\sqrt{E^2-1}}}$
and $\tau'$ as
$\frac{\tau'}{M} = -\exp{\left( \frac{\tau}{4ME} \right) }=\allowbreak
=-
\sqrt{\frac{2M}{r}} \frac{\sqrt{1-\frac{r}{2M}}} {\sqrt{2E^2-
1+\frac{2M}{r}+ 2E\sqrt{E^2-1+\frac{2M}{r}}} }
\exp{\left(
\frac{t}{4M} +\frac{r}{4ME} \sqrt{E^2-1+ \frac{2M}{r}}\right)}
\left[ \frac{r}{M} \left( \sqrt{E^2-1} \sqrt{E^2-1+\frac{2M}{r}}+
E^2-1+\right.\right.\\
\left.\left.+\frac{M}{r} \right)\right]^{\frac{2E^2-1}{4E \sqrt{E^2-1}}}$.
These
transformations are valid for the region with
${\mathcal t}'\leq 0$ and
$\tau'\leq
0$. It is a region with $r\leq 2M$ and
$-\infty<t<\infty$.
Note that the coordinate transformations in this region give
$\left( \frac{1- \frac{r}{2M}}{2E^2-1+ \frac{2M}{r}+ 2E \sqrt{E^2-1+
\frac{2M}{r}}} \right) \frac{2M}{r}\,\times \allowbreak
\times \exp{\left( \frac{r}{2ME}
\sqrt{E^2-1+ \frac{2M}{r}} \right)}
\left[ \frac{r}{M} \left( E^2-1 +
\frac{M}{r}+
\sqrt{E^2-1} \sqrt{E^2-1+ \frac{2M}{r}}\right)
\right]^{\frac{2E^2-1}{2E \sqrt{E^2-1}}}= \frac{{\mathcal t}'}{M}
\frac{\tau'}{M}$.
But all
this has been automatically incorporated
into Eqs.~(\ref{eqn:metricE>1}) and (\ref{rimplicitEoo})
so once again there is no further concern on that.

Furthermore, from Eq.~(\ref{rimplicitEoo}) we see that the event
horizon at $r=2M$ has two solutions, ${\mathcal t}'=0$ and $\tau'=0$
which are null surfaces represented
by straight lines. The true curvature singularity at $r=0$ has
two solutions $\frac{{\mathcal t}'}{M}\frac{\tau'}{M}=1$, i.e., two
spacelike hyperbolae. Implicit in the construction, there is a
wormhole, or Einstein-Rosen bridge, topology, with its throat
expanding and contracting.  The dynamic wormhole is non traversable,
but it spatially connects region I to region III through regions II
and IV.  Regions I and III are two asympotically flat regions causally
separated, region II is the black hole region, and region IV is the
white hole region of the spacetime.

\begin{figure}[h]
\centering
\subfigure[]
{\label{fig:diagramEgt1a}
\includegraphics[scale=0.3]{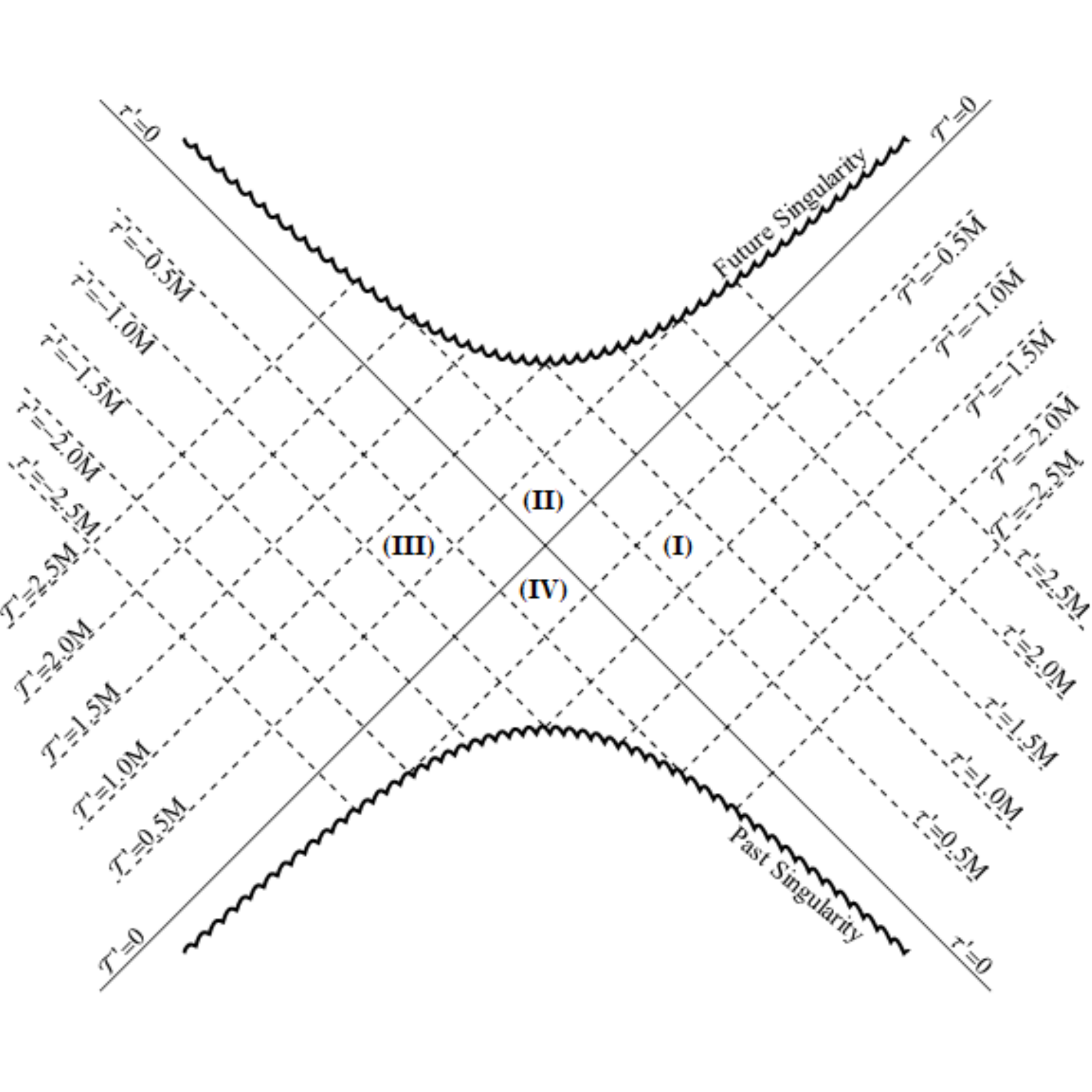}}
\qquad
\subfigure[]
{\label{fig:diagramEgt1b}
\includegraphics[scale=0.3]{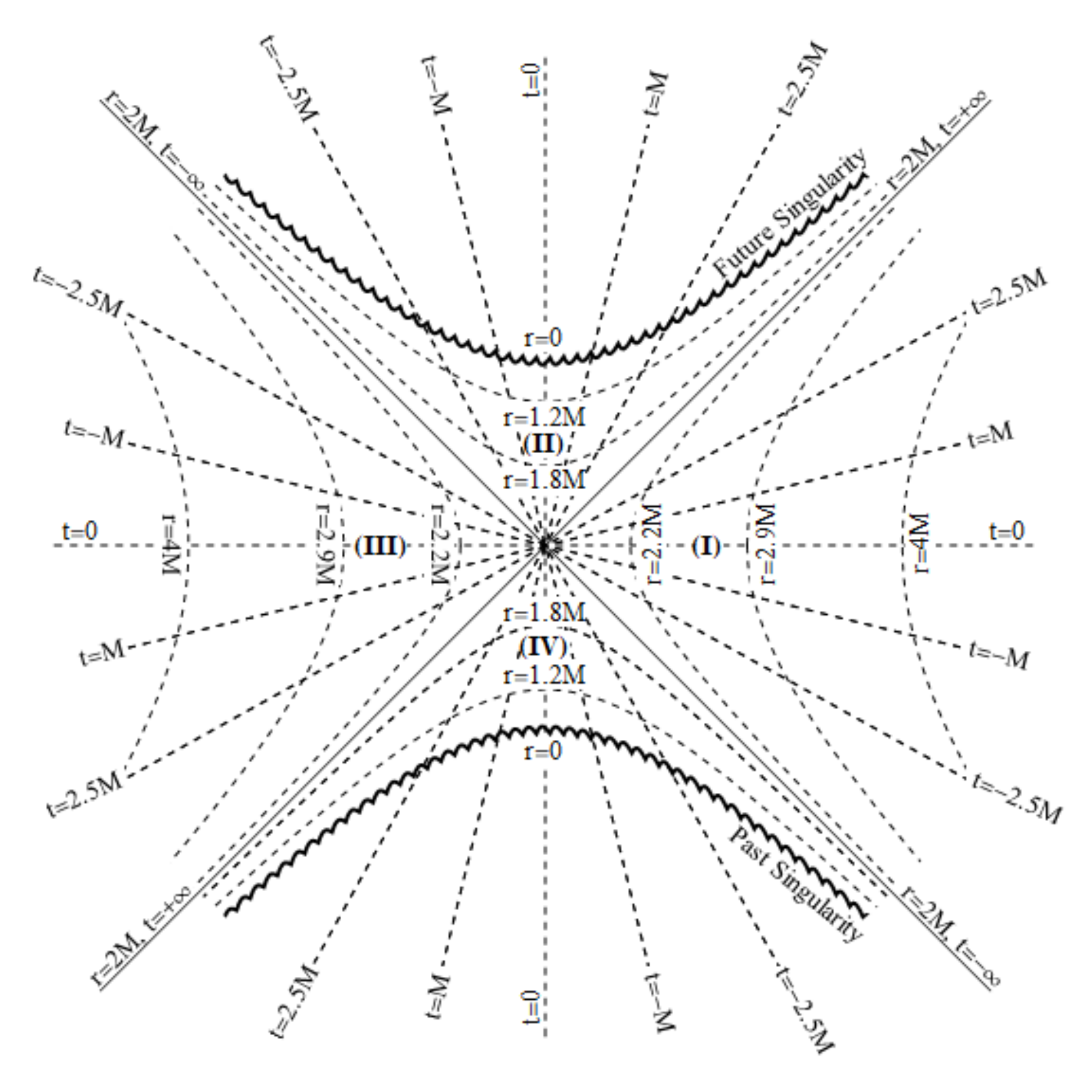}}
\caption{The maximal analytical extension of the Schwarzschild metric
for the parameter $E$ generic obeying $E>1$ in the plane $({\mathcal
t}',\tau')$ is shown in a diagram with two different descriptions,
(a) and (b). In (a) typical values for lines of constant ${\mathcal
t}'$ and constant $\tau'$ are displayed.  In (b) typical values for
lines of constant $t$ and constant $r$ are displayed.  The diagram,
both in (a) and in (b), represents a
a spacetime with a wormhole, not shown, that forms
out of a singularity in the white hole region, i.e., region IV, and
finishes at the black hole region and its singularity, i.e., region
II, connecting the two separated asymptotically flat spacetimes,
regions I and III.}
\label{fig:diagramEgt1}
\end{figure}

Eqs.~(\ref{eqn:metricE>1}) and~(\ref{rimplicitEoo}) together with the
corresponding interpretation give the maximal extension of the
Schwarzschild metric for $E>1$, in the coordinates $({\mathcal
t}',\tau',\theta,\phi)$.  Since $1<E<\infty$ this is a family of
extensions, characterized by one parameter, the parameter $E$.  It is
a one-parameter family of extensions.  The two-dimensional part
$({\mathcal t}',\tau')$ of the coordinate system $({\mathcal
t}',\tau',\theta,\phi)$ is shown in Figure~\ref{fig:diagramEgt1}, both
for lines of constant ${\mathcal t}'$ and constant $\tau'$ in part (a)
of the figure, and for lines of constant $t$ and constant $r$ in part
(b) of the figure, conjointly with the labeling of regions I, II, III,
IV, needed to cover it.

\newpage

It is also worth discussing the normals to
the ${\mathcal t}'={\rm constant}$ and
$\tau'={\rm constant}$ hypersurfaces.
From
Eq.~(\ref{eqn:metricE>1}) one finds that
the covariant metric
has components
$g_{{\mathcal t'} {\mathcal t'}}=\frac{4}{M^2} \left(
\frac{ \left(2E^2-1+ \frac{2M}{r}+ 2E
\sqrt{E^2-1+\frac{2M}{r}}\right)^2}{ E^2-1+\frac{2M}{r}} \right)
\exp{\left( -\frac{r}{ME} \sqrt{E^2-1+ \frac{2M}{r}} \right)} \times\allowbreak
\times
\left( \frac{\frac{M}{r}}{E^2-1+ \frac{M}{r}+
\sqrt{E^2-1}\sqrt{E^2-1+\frac{2M}{r}}} \right)
^{\frac{2E^2-1}{E \sqrt{E^2-1}}}
\tau'^2
$,
$g_{\tau' \tau'}= \frac{4}{M^2} \left( \frac{ \left( 2E^2-1+
\frac{2M}{r}+ 2E \sqrt{E^2-1+\frac{2M}{r}}\right)^2}{
E^2-1+\frac{2M}{r}} \right)
\exp{\left( -\frac{r}{ME} \sqrt{E^2-1+
\frac{2M}{r}} \right)}
\times \allowbreak
\times
\left( \frac{\frac{M}{r}}{E^2-1+ \frac{M}{r}+
\sqrt{E^2-1}\sqrt{E^2-1+\frac{2M}{r}}} \right)
^{\frac{2E^2-1}{E \sqrt{E^2-1}}}
{\mathcal t}'^2$,
$g_{{\mathcal t'} \tau'}=g_{\tau' {\mathcal t'}}=- 4
\left( \frac{ 2E^2-1+ \frac{2M}{r}+ 2E
\sqrt{E^2-1+\frac{2M}{r}}}{ E^2-1+\frac{2M}{r}} \right)
\left( 2E^2-1+ \frac{2M}{r} \right)\times \allowbreak
\times
\exp{\left( -\frac{r}{2ME} \sqrt{E^2-1+ \frac{2M}{r}} \right)}
\left( \frac{\frac{M}{r}}{E^2-1+ \frac{M}{r}+
\sqrt{E^2-1}\sqrt{E^2-1+\frac{2M}{r}}} \right)
^{\frac{2E^2-1}{2E \sqrt{E^2-1}}}
$,
$g_{\theta \theta}=r^2$,
$g_{\phi \phi}=r^2\sin^2{\theta}$.
The contravariant components of the metric
can be calculated to be
$g^{{\mathcal t}' {\mathcal t}'}= - \frac{{\mathcal t}'^2}{16 M^2 E^2}$,
$g^{{\tau}' {\tau}'}=- \frac{\tau'^2}{16 M^2 E^2}$,
$g^{{\mathcal t}' \tau'}=g^{\tau' {\mathcal t}'}=\allowbreak
=-
\frac{1}{16E^2}
\frac{1}{
2E^2-1+ \frac{2M}{r}+ 2E \sqrt{E^2-1+ \frac{2M}{r}}
}
\left(2E^2-1+ \frac{2M}{r} \right)
\left[
\frac{r}{M}\left(E^2-1+ \frac{M}{r}+ \sqrt{E^2-1} \sqrt{E^2-1+
\frac{2M}{r}}\right) \right]^{\frac{2E^2-1}{2E \sqrt{E^2-1}}}
\times
\\
\times \exp{\left( \frac{r}{2ME} \sqrt{E^2-1+ \frac{2M}{r}} \right)}$,
$g^{\theta \theta}=\frac{1}{r^2}$,
$g^{\phi \phi}=\frac{1}{r^2\sin^2{\theta}}$.
The normals $n_a$ to the ${\mathcal t}'={\rm constant}$ and
$\tau'={\rm constant}$ hypersurfaces are
${n^{\mathcal t'}}_a=(1,0,0,0)$ and
${n^{\tau'}}_a=(0,1,0,0)$, respectively, where the superscripts
${\mathcal t'}$ and $\tau'$ in this context
are not indices, they simply
label the respective normal. 
Their contravariant components are, respectively,
$n^{{\mathcal t'}\hskip0.005cm a}
=(g^{{\mathcal t}' {\mathcal t}'}, g^{{\mathcal
t}'\tau'},0,0)$ and $n^{\tau'\hskip0.005cm a}
=(g^{\tau'{\mathcal t}'},g^{\tau'
\tau'},0,0)$, awkward writing them explicitly due to the long
expression for $g^{{\mathcal t}'\tau'}$.
The norms are
then $
{n^{\mathcal t'}}_a n^{{\mathcal t'}\hskip0.005cm a}
=-
\frac{{\mathcal t}'^2}{16 M^2 E^2}$ and
$
{n^{\tau'}}_a n^{{\tau'}\hskip0.005cm a} 
=- \frac{\tau'^2}{16 M^2 E^2}$, respectively.
Thus, clearly, the normals to the ${\mathcal t'}={\rm constant}$ and
$\tau'={\rm constant}$ hypersurfaces are timelike, and so ${\mathcal
t}'$ and $\tau'$ are timelike coordinates, and the corresponding
hypersurfaces are spacelike, only in a measure zero are they null,
when ${\mathcal t}'=0$ and $\tau'=0$, respectively.

\vfill


\section{Maximal analytic extension for $E=1$ as the
lower limit of $E>1$}
\label{E=1m}

To build the maximal analytic extension for $E=1$, we take the $E\to
1$ limit from the $E>1$ case.
Using 
$\ln\Big[ \left(2\sqrt{E^2-1}\sqrt{\frac{r}{2M}}\right)+1\Big]
=2\sqrt{E^2-1}\sqrt{\frac{r}{2M}}$ in this limit, we find 
that the coordinates
${\mathcal t}$ and $\tau$ of Eqs.~(\ref{eqn:closedrel2gt1}) and
(\ref{eqn:closedrel1gt1}) become
\begin{align}
\label{tE1}
{\mathcal t}&= t- 4M \sqrt{\frac{r}{2M}}+ 2M \ln{\left|
\frac{\sqrt{\frac{r}{2M}}+1}{\sqrt{\frac{r}{2M}}-1} \right|}\,,\\
\label{tauE1}
\tau&= t+ 4M \sqrt{\frac{r}{2M}}-
2M \ln{\left|
\frac{\sqrt{\frac{r}{2M}}+1}{\sqrt{\frac{r}{2M}}-1} \right|}
\,.
\end{align}
The line element given in  Eq.~(\ref{eqn:genmetric})
is then in this $E=1$ limit given by
\begin{equation}
\label{metricuvEeq1}
\mathrm{d}s^2= -\frac14
\frac{
\left( 1-\frac{2M}{r}\right)}{\frac{2M}{r}}
\left[ -\left(1-
\frac{2M}{r} \right) (\mathrm{d}{\mathcal t}^2+ \mathrm{d}\tau^2)+ 2
\left(1+ \frac{2M}{r} \right) \mathrm{d} {\mathcal t}\, \mathrm{d}\tau
\right]+ r^2( \mathrm{d}\theta^2+ \sin^2{\theta}\,
\mathrm{d}\phi^2)\,,
\end{equation}
with $r=r({\mathcal t},\tau)$ being
obtained via Eq.~(\ref{rimplicitE>1})
in the $E=1$ limit, or through Eqs.~(\ref{tE1})
and (\ref{tauE1}), i.e.,
\begin{equation}
\label{rimplicitE1}
4M \sqrt{\dfrac{r}{2M}}-
2M \ln{\left|
\frac{\sqrt{\frac{r}{2M}}+1}{\sqrt{\frac{r}{2M}}-1}
\right|}=\frac12\left(-{\mathcal t}+\tau\right)\,.
\end{equation}

Again, as in Eq.~(\ref{eqn:genmetric}),
the line element given in Eq.~(\ref{metricuvEeq1})
is still degenerate at
$r=2M$. So, to extend it past 
$r=2M$
we again make use of
maximal extended coordinates,
${\mathcal t}'$ and $\tau'$, 
defined
as $\frac{{\mathcal t}'}{M}= -\exp{\left(
-\frac{{\mathcal t}}{4M} \right)}$
and $\frac{\tau'}{M}= \exp{\left( \frac{\tau}{4M} \right)}$,
which by either taking directly the limit $E=1$ in 
Eqs.~(\ref{eqn:coordtransfrho2gt1})
and (\ref{eqn:coordtransftau2gt1}), respectively,
or using 
Eqs.~(\ref{tE1})
and (\ref{tauE1}),
yields for $r>2M$,
\begin{align}
\label{tdash1}
\frac{{\mathcal t}'}{M}&= -\exp{\left(
-\frac{{\mathcal t}}{4M} \right)}\,,\quad {\rm i.e.,}
\quad
\frac{{\mathcal t}'}{M}= -\sqrt{
\frac
{\sqrt{\frac{r}{2M}}-1}
{\sqrt{\frac{r}{2M}}+1}
}
\exp{\left( - \frac{t}{4M}+
\sqrt{\frac{r}{2M}} \right)}\,,\\
\label{taudash1}
\frac{\tau'}{M}&=\quad \exp{\left( \;\frac{\tau}{4M} \right)}\,,
\;\;\,\quad {\rm i.e.,}\quad
\frac{\tau'}{M}= \quad
\sqrt{
\frac
{\sqrt{\frac{r}{2M}}-1}
{\sqrt{\frac{r}{2M}}+1}
}
\exp{\left(\;\frac{t}{4M}+ \sqrt{\frac{r}{2M}} \right)}\,,
\end{align}
respectively.
Through the $E=1$ limit of Eq.~(\ref{eqn:metricE>1}),
or putting ${\mathcal t}'$  and $\tau'$ given in
Eqs.~(\ref{tdash1})
and (\ref{taudash1}),
respectively, into the line element
Eq.~(\ref{metricuvEeq1}),
one finds that the new $E=1$ line element
in coordinates $({\mathcal t}',\tau',\theta,\phi)$
is given by
\begin{equation}
\label{metricu'v'Eeq1}
\begin{split}
\mathrm{d}s^2= &-4
\frac{\left( 1+ \sqrt{\frac{2M}{r}} \right)^2}
{\frac{2M}{r}}
\exp{\left( -2 \sqrt{\frac{r}{2M}} \right)}
\left[ -\frac{1}{M^2} \left( 1+
\sqrt{\frac{2M}{r}} \right)^2 \exp{\left( -2 \sqrt{\frac{r}{2M}}
\right)} (\tau'^2\, \mathrm{d}{\mathcal t}'^2 +{\mathcal t}'^2\,
\mathrm{d}\tau'^2)+\right.
\\
&\left. + 2\left(1+ \frac{2M}{r} \right) \mathrm{d}
{\mathcal t}'\, \mathrm{d} \tau' \right]+
r^2 (\mathrm{d}\theta^2+ \sin^2{\theta}\, \mathrm{d}\phi^2) \,,
\end{split}
\end{equation}
with $r=r({\mathcal
t}',\tau')$ given implicitly,  see Eq.~(\ref{rimplicitEoo})
in the $E=1$ limit,
or directly through Eqs.~(\ref{tdash1}) and (\ref{taudash1}), by
\begin{equation}
\label{rimplicitu'v'Eeq1}
\frac
{\sqrt{\frac{r}{2M}}-1}
{\sqrt{\frac{r}{2M}}+1}
\exp{\left( 2
\sqrt{\frac{r}{2M}} \right)} =-\frac{{\mathcal
t}'}{M}\frac{\tau'}{M}\,.
\end{equation}
All of this is done so that ${\mathcal t}'$ and $\tau'$ have ranges
$-\infty<{\mathcal t}'<\infty$ and $-\infty<\tau'<\infty$, which
Eqs.~(\ref{metricu'v'Eeq1}) and (\ref
{rimplicitu'v'Eeq1}) permit.
To obtain Eqs.~(\ref{metricu'v'Eeq1}) and (\ref{rimplicitu'v'Eeq1})
directly from the $E\to1$ limit of Eqs.~(\ref{eqn:metricE>1})
and~(\ref{rimplicitEoo}), respectively, see
the Appendix.
Several
properties are again worth mentioning.

In terms of the coordinates $({\mathcal t},\tau)$, or
$(t,r)$, the coordinate
transformations that yield the maximal extended coordinates
$({\mathcal t}',\tau')$
with infinite ranges have to be broadened, resulting
in the existence of four regions, regions I, II, III, and IV.
Region I is the region where the transformations
Eqs.~(\ref{tdash1}) and~(\ref{taudash1})
hold, i.e., it is a region with ${\mathcal t}'\leq 0$ and $\tau'\geq
0$. It is a region with $r\geq 2M$ and $-\infty<t<\infty$.
Of course, in this region
Eqs.~(\ref{metricu'v'Eeq1}) and~(\ref{rimplicitu'v'Eeq1}) hold.
Region II, a region for which $r\leq 2M$, gets a different set of
coordinate transformations.  In this $r\leq 2M$ region, due to the
moduli appearing in Eqs.~(\ref{tdash1})
and~(\ref{taudash1}) and the change of sign in
Eq.~(\ref{metricuvEeq1}), one defines instead
${\mathcal t}'$ as
$ \frac{{\mathcal t}'}{M}=+
\exp{\left(-\frac{{\mathcal
t}}{4M}\right)}=\\
=+\sqrt{
\frac
{1-\sqrt{\frac{r}{2M}}}
{1+\sqrt{\frac{r}{2M}}}
}
\exp{\left( - \frac{t}{4M}+
\sqrt{\frac{r}{2M}} \right)}$
and $\tau'$ as
$\frac{\tau'}{M} = \exp{\left( \frac{\tau}{4M} \right) }=
\sqrt{
\frac
{1-\sqrt{\frac{r}{2M}}}
{1+\sqrt{\frac{r}{2M}}}
}
\exp{\left(\;\frac{t}{4M}+ \sqrt{\frac{r}{2M}} \right)}$.
These transformations are valid for ${\mathcal t}'\geq 0$ and
$\tau'\geq 0$.
It is a region with $r\leq2M$ and $-\infty<t<\infty$.
Note that the coordinate transformations in this region give
$\frac
{1-\sqrt{\frac{r}{2M}}}
{1+\sqrt{\frac{r}{2M}}}
\exp{\left( 2
\sqrt{\frac{r}{2M}} \right)} =\frac{{\mathcal
t}'}{M}\frac{\tau'}{M}$
But all
this has been automatically incorporated
into Eqs.~(\ref{metricu'v'Eeq1}) and (\ref{rimplicitu'v'Eeq1})
so there is no further concern on that.
Region III is another $r\geq 2M$ region.
Now one defines ${\mathcal t}'$
as
$ \frac{{\mathcal t}'}{M}=
\exp{\left(-\frac{{\mathcal
t}}{4M}\right)}=\sqrt{
\frac
{\sqrt{\frac{r}{2M}}-1}
{\sqrt{\frac{r}{2M}}+1}
}
\exp{\left( - \frac{t}{4M}+
\sqrt{\frac{r}{2M}} \right)}$
and $\tau'$ as
$\frac{\tau'}{M} = -\exp{\left( \frac{\tau}{4M} \right) }=
-\sqrt{
\frac
{\sqrt{\frac{r}{2M}}-1}
{\sqrt{\frac{r}{2M}}-1}
}
\exp{\left(\;\frac{t}{4M}+ \sqrt{\frac{r}{2M}} \right)}$.
These transformations are valid for the region with ${\mathcal t}'\geq
0$ and $\tau'\leq 0$. It is a region with $r\geq 2M$ and
$-\infty<t<\infty$.  Note that the coordinate transformations in this
region give
$\frac
{\sqrt{\frac{r}{2M}}-1}
{\sqrt{\frac{r}{2M}}+1}
\exp{\left( 2
\sqrt{\frac{r}{2M}} \right)} =-\frac{{\mathcal
t}'}{M}\frac{\tau'}{M}$
But all
this has been automatically incorporated
into Eqs.~(\ref{metricu'v'Eeq1}) and (\ref{rimplicitu'v'Eeq1})
so again there is no further concern on that.
\noindent
Region IV is another region with $r\leq 2M$.
Now, one defines ${\mathcal t}'$
as
$ \frac{{\mathcal t}'}{M}=-
\exp{\left(-\frac{{\mathcal
t}}{4M}\right)}=-\sqrt{
\frac
{1-\sqrt{\frac{r}{2M}}}
{1+\sqrt{\frac{r}{2M}}}
}
\exp{\left( - \frac{t}{4M}+
\sqrt{\frac{r}{2M}} \right)}$
and $\tau'$ as
$\frac{\tau'}{M} =- \exp{\left( \frac{\tau}{4M} \right) }=
-\sqrt{
\frac
{1-\sqrt{\frac{r}{2M}}}
{1+\sqrt{\frac{r}{2M}}}
}
\exp{\left(\;\frac{t}{4M}+ \sqrt{\frac{r}{2M}} \right)}$.
These
transformations are valid for the region with
${\mathcal t}'\leq 0$ and
$\tau'\leq
0$. It is a region with $r\leq 2M$ and
$-\infty<t<\infty$.
Note that the coordinate transformations in this region give
$\frac
{1-\sqrt{\frac{r}{2M}}}
{1+\sqrt{\frac{r}{2M}}}
\exp{\left( 2
\sqrt{\frac{r}{2M}} \right)} =\frac{{\mathcal
t}'}{M}\frac{\tau'}{M}$.
But all
this has been automatically incorporated
into Eqs.~(\ref{metricu'v'Eeq1}) and (\ref{rimplicitu'v'Eeq1})
so once again there is no further concern on that.

Furthermore, from Eq.~(\ref{rimplicitu'v'Eeq1}) we see that the event
horizon at $r=2M$ has two solutions, ${\mathcal t}'=0$ and $\tau'=0$
which are null surfaces represented
by straight lines. The true curvature singularity at $r=0$ has
two solutions $\frac{{\mathcal t}'}{M}\frac{\tau'}{M}=1$, i.e., two
spacelike hyperbolae. Implicit in the construction, there is a
wormhole, or Einstein-Rosen bridge, topology, with its throat
expanding and contracting.  The dynamic wormhole is non traversable,
but it spatially connects region I to region III through regions II
and IV.  Regions I and III are two asympotically flat regions causally
separated, region II is the black hole region, and region IV is the
white hole region of the spacetime.

\begin{figure}[h]
\centering
\subfigure[]{
\includegraphics[scale=0.3]{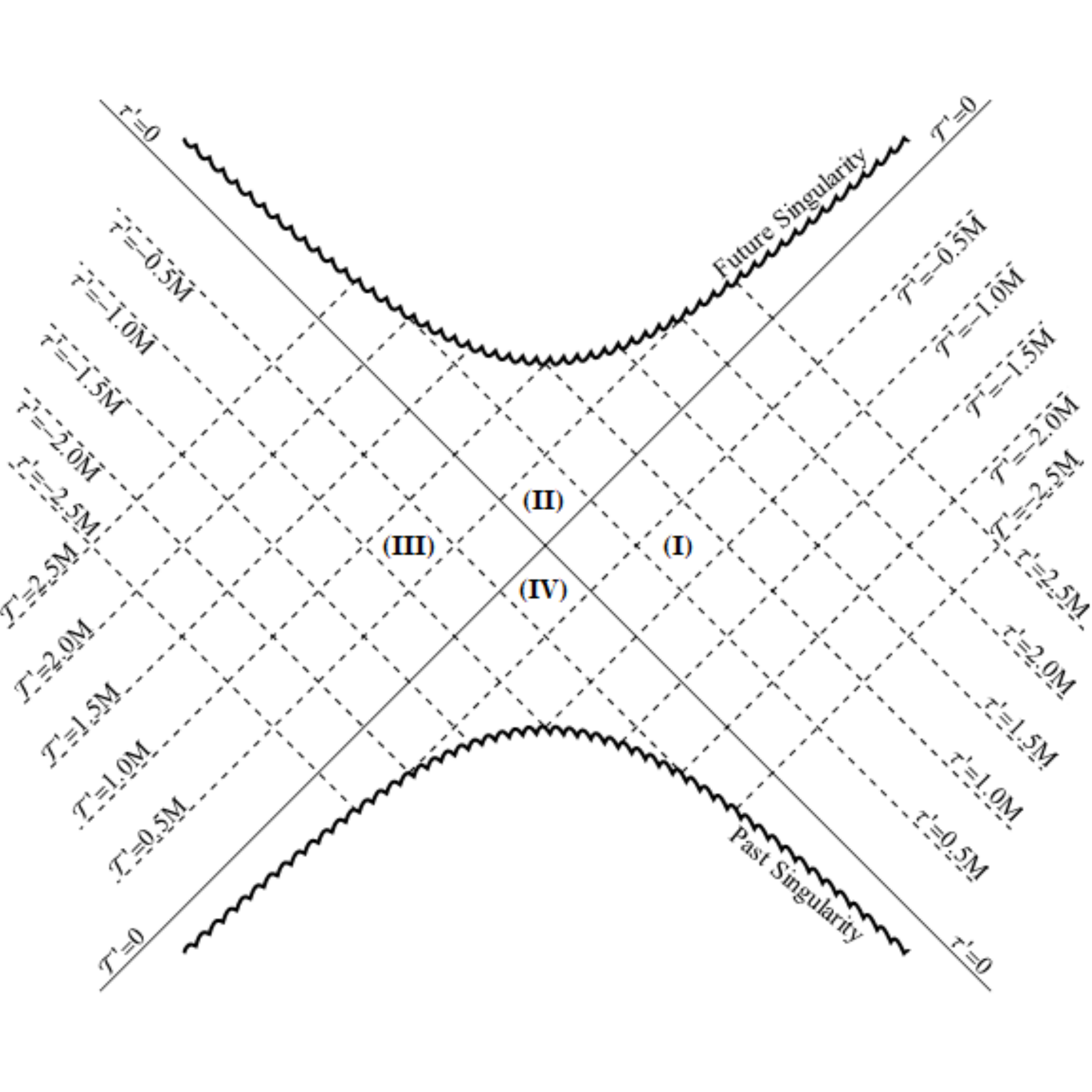}
\label{fig:diagramEeq1a}
}\qquad
\subfigure[]{
\includegraphics[scale=0.3]{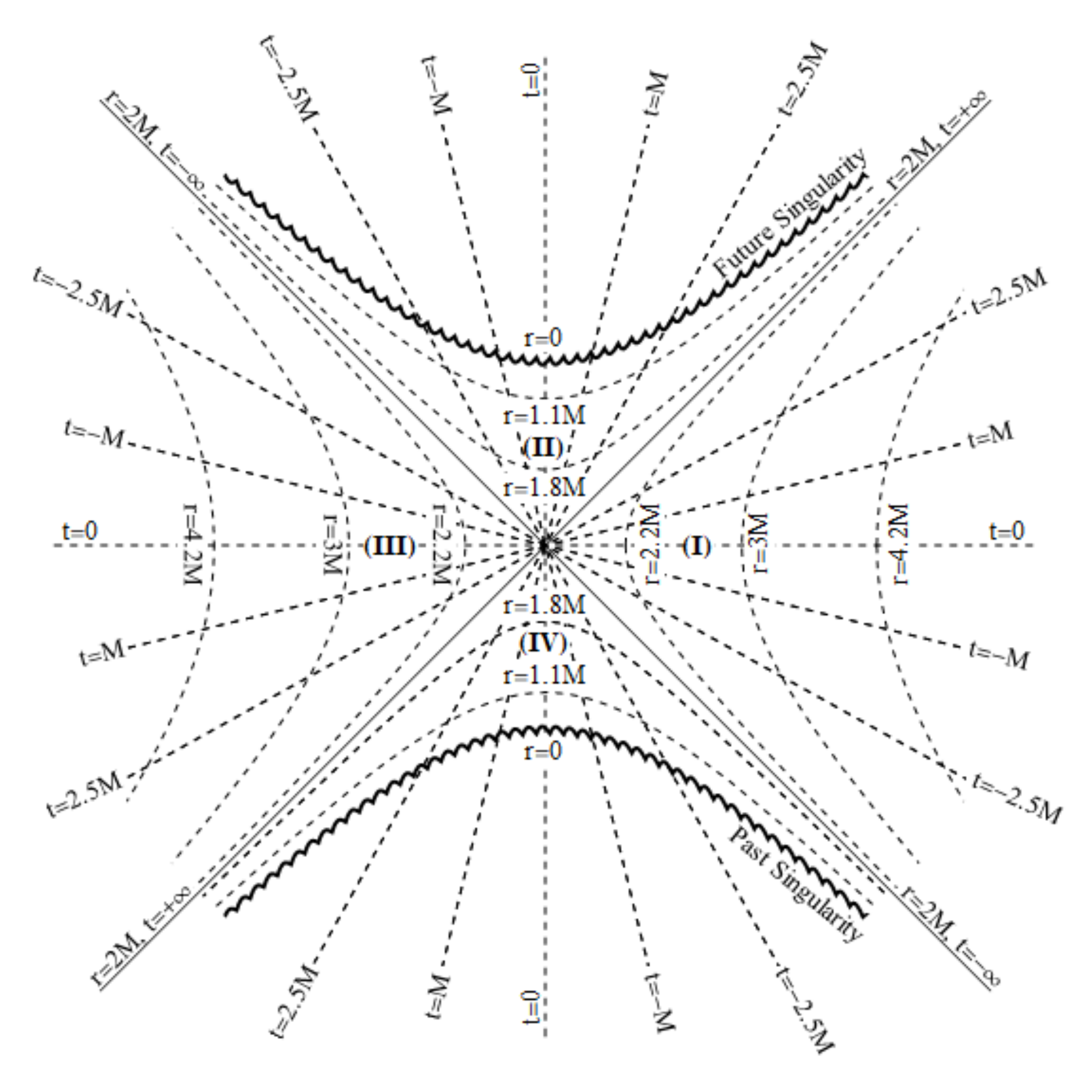}
\label{fig:diagramEeq1b}
}
\caption{The maximal analytical extension of the Schwarzschild metric
for the parameter $E=1$ in the plane $({\mathcal t}',\tau')$ is shown
in a diagram with two different descriptions, (a) and (b). In (a)
typical values for lines of constant ${\mathcal t}'$ and constant
$\tau'$ are displayed.  In (b) typical values for lines of constant $t$
and constant $r$ are displayed.  The diagram, both in (a) and in
(b), represents  a
a spacetime with a wormhole, not shown, that forms out of a singularity
in the white hole region, i.e., region IV, and finishes at the black
hole region and its singularity, i.e., region II, connecting the two
separated asymptotically flat spacetimes, regions I and III.  The
$E=1$ diagram is very similar to the $E>1$ generic case diagram, see
Figure~\ref{fig:diagramEgt1}, as it is expected for a maximal
extension of the Schwarzschild spacetime, in particular for those
extensions within the same family.
}
\label{fig:diagramEeq1}
\end{figure}

Eqs.~(\ref{metricu'v'Eeq1}) and (\ref{rimplicitu'v'Eeq1}) together
with the corresponding interpretation give the maximal extension of
the Schwarzschild metric for $E=1$, in the coordinates $({\mathcal
t}',\tau',\theta,\phi)$. The two-dimensional part
$({\mathcal t}',\tau')$ of the coordinate system $({\mathcal
t}',\tau',\theta,\phi)$ is shown in Figure~\ref{fig:diagramEeq1}, both
for lines of constant ${\mathcal t}'$ and constant $\tau'$ in part (a)
of the figure, and for lines of constant $t$ and constant $r$ in part
(b) of the figure, conjointly with the labeling of regions I, II, III,
IV, needed to cover it.

It is also worth discussing the normals to the
${\mathcal t}'={\rm constant}$ and
$\tau'={\rm constant}$ hypersurfaces.
From
Eq.~(\ref{metricu'v'Eeq1}) one finds that
the metric
has covariant components
$g_{{\mathcal t'} {\mathcal t'}}=\frac{4}{M^2}
\frac{
\left(1+
\sqrt{\frac{2M}{r}} \right)^4}{\frac{2M}{r}} \exp{\left( -4
\sqrt{\frac{r}{2M}} \right)} \tau'^2 $,
$g_{\tau' \tau'}= \frac{4}{M^2}
\frac{
\left(1+
\sqrt{\frac{2M}{r}} \right)^4}{\frac{2M}{r}}
\times
\\
\times
\exp{\left( -4
\sqrt{\frac{r}{2M}} \right)}
{\mathcal t}'^2$,
$g_{{\mathcal t'} \tau'}=g_{\tau' {\mathcal t'}}=
-4
\frac{
\left( 1+  \sqrt{\frac{2M}{r}} \right)^2}{\frac{2M}{r}}
\left( 1+ \frac{2M}{r} \right)
\exp{\left( -2
\sqrt{\frac{r}{2M}} \right)}$,
$g_{\theta \theta}=r^2$,
$g_{\phi \phi}=r^2\sin^2{\theta}$.
The contravariant components of the metric
can be calculated to be
$g^{{\mathcal t}' {\mathcal t}'}= - \frac{{\mathcal t}'^2}{16 M^2}$,
$g^{{\tau}' {\tau}'}= -\frac{\tau'^2}{16M^2}$,
$g^{{\mathcal t}' \tau'}=g^{\tau' {\mathcal t}'}=
-\frac{1}{16}\frac{1+\frac{2M}{r}}
{\left(1+ \sqrt{\frac{2M}{r}} \right)^{2}} \times \\ \times
\exp{\left( 2 \sqrt{\frac{r}{2M}}
\right)}$,
$g^{\theta \theta}=\frac{1}{r^2}$,
$g^{\phi \phi}=\frac{1}{r^2\sin^2{\theta}}$.
The normals $n_a$ to the ${\mathcal t}'={\rm constant}$ and
$\tau'={\rm constant}$ hypersurfaces are
${n^{\mathcal t'}}_a=(1,0,0,0)$ and
${n^{\tau'}}_a=(0,1,0,0)$, respectively, where the superscripts
${\mathcal t'}$ and $\tau'$ in this context are not indices, they simply
label the respective normal. 
Their contravariant components are
$n^{{\mathcal t'}\hskip0.005cm a}
=(g^{{\mathcal t}' {\mathcal t}'},
g^{{\mathcal t}'\tau'},0,0) =\allowbreak {=
(- \frac{{\mathcal t}'^2}{16 M^2},
-\frac{1}{16}\frac{1+\frac{2M}{r}}
{\left(1+ \sqrt{\frac{2M}{r}} \right)^{2}}
\exp{\left( 2 \sqrt{\frac{r}{2M}}
\right)}}
,0,0)$
and
$n^{\tau'\hskip0.005cm a}=(g^{\tau'{\mathcal t}'},g^{\tau' \tau'},0,0)
=
(-\frac{1}{16}\frac{1+\frac{2M}{r}}
{\left(1+ \sqrt{\frac{2M}{r}} \right)^{2}}
\exp{\left( 2 \sqrt{\frac{r}{2M}}
\right)},-\frac{\tau'^2}{16M^2},0,0)$, respectively.
The norms are
then $
{n^{\mathcal t'}}_a n^{{\mathcal t'}\hskip0.005cm a}
=-
\frac{{\mathcal t}'^2}{16 M^2}$ and
$
{n^{\tau'}}_a n^{{\tau'}\hskip0.005cm a} 
=- \frac{\tau'^2}{16 M^2}$, respectively.
Thus, clearly, the normals to the ${\mathcal t'}={\rm constant}$ and 
$\tau'={\rm constant}$ hypersurfaces
are timelike, and so 
${\mathcal t}'$ and 
$\tau'$ are timelike coordinates, and the corresponding hypersurfaces
are spacelike, only in a measure zero are they  null, 
when ${\mathcal t}'=0$ and $\tau'=0$, respectively.
The metric components and the normals
can also be found from the $E>1$ case in the
$E=1$ limit.


\vskip 1.5cm
\section{Maximal analytic extension for $E= \infty$ as the upper limit
of $E>1$: The Kruskal-Szekeres maximal extension of the Schwarzschild
metric}
\label{E=oom}

To build the maximal analytic extension for $E=\infty$,
we take the $E\to \infty$ limit from the $E>1$ generic case.
We will see that this limit is
the Kruskal-Szekeres maximal analytic extension.
Taking a
redefinition of the coordinates $\tau$ and ${\mathcal t}$ of
Eqs.~(\ref{eqn:closedrel2gt1}) and (\ref{eqn:closedrel1gt1}) 
to coordinates $u$ and $v$, respectively, 
we find that these become
\begin{align}
\label{uedd}
u&\equiv
\lim_{E\to \infty} \frac{{\mathcal t}}{E}\,, \quad {\rm i.e.,}
\quad\quad
u=t-r-2M \ln{\left| \frac{r}{2M}-1 \right|}\,,
\\
\label{vedd}
v&\equiv\lim_{E\to \infty} \frac{\tau}{E}\,, \quad {\rm i.e.,}
\quad\quad
v=t+r+2M \ln{\left| \frac{r}{2M}-1 \right|}\,.
\end{align}
The line element given in  Eq.~(\ref{eqn:genmetric})
is then in this limit 
\begin{equation}
\label{metricuv}
\mathrm{d}s^2=-
\left(1-\frac{2M}{r}\right)
dudv+
r^2( \mathrm{d}\theta^2+
\sin^2{\theta}\, \mathrm{d}\phi^2)\,.
\end{equation}
with $r=r(u,v)$ being
obtained directly via Eq.~(\ref{rimplicitE>1})
in the $E\to\infty$ limit, or through Eqs.~(\ref{uedd})
and (\ref{vedd}), i.e.,
\begin{equation}
\label{rimplicitEooueddvedd}
r+2M \ln{\left| \frac{r}{2M}-1 \right|}=\frac12\left(-u+v\right)\,.
\end{equation}

Again, the line element given in Eq.~(\ref{metricuv})
is still degenerate at
$r=2M$. So, to extend it past 
$r=2M$,
we make use of the maximal extended timelike
coordinates ${\mathcal t}'$ and $\tau'$ 
defined through
$\frac{{\mathcal t}'}{M}= -\exp{\left( -
\frac{{\mathcal t}}{4ME}\right)}$
and $\frac{\tau'}{M}= \exp{\left( \;\frac{\tau}{4ME} \right)}$,
which in this limit $E\to\infty$ are redefined to
maximal extended coordinates, 
$u'$ and $v'$, respectively,
obtained directly via
Eqs.~(\ref{eqn:coordtransfrho2gt1})
and~(\ref{eqn:coordtransftau2gt1})
in the $E\to\infty$ limit, or using 
Eqs.~(\ref{uedd}) and (\ref{vedd}), to find
\begin{align}
\label{udashed}
u'&=\lim_{E\to \infty} {\mathcal t'}\,, \quad {\rm i.e.,}
\quad\quad
\frac{u'}{M}=-\exp{\left(-\frac{u}{4M}\right)}\,,\quad\quad
{\rm i.e.,} \quad\quad
\frac{u'}{M}=-\sqrt{\frac{r}{2M}-1}\,
\exp{\left( -\frac{t}{4M}+\frac{r}{4M}\right)}\,,
\\
\label{vdashed}
v'&=\lim_{E\to \infty} \tau'\,, \quad {\rm i.e.,} \quad\quad
\frac{v'}{M}=\quad\exp{\left(\frac{v}{4M}\right)}\,,
\quad\quad\;\;\; {\rm i.e.,} \quad\quad
\frac{v'}{M}=\quad \sqrt{\frac{r}{2M}-1}\,
\exp{\left( \frac{t}{4M}+\frac{r}{4M}\right)}\,.
\end{align}
\hfill
\vskip 4cm
\noindent
Then, the line element of (\ref{eqn:metricE>1})
in the $E\to\infty$ limit, or through Eq.~(\ref{metricuv})
together with Eqs.~(\ref{udashed}) and (\ref{vdashed}),
yields the new line element
\begin{equation}
\label{metricu'v'}
\mathrm{d}s^2=-\frac{32M}{r} \exp{\left(-\frac{r}{2M}\right)}
\mathrm{d}u'\, \mathrm{d}v'+ r^2( \mathrm{d}\theta^2+
\sin^2{\theta}\, \mathrm{d}\phi^2)\,,
\end{equation}
with $r=r(u',v')$ given implicitly, see Eq.~(\ref{rimplicitEoo})
in the $E\to\infty$ limit,
or directly through Eqs.~(\ref{udashed}) and (\ref{vdashed}),
by
\begin{equation}
\label{rimplicitu'v'}
\left(\frac{r}{2M}-1\right)
\exp{\left(\frac{r}{2M}\right)}=-\frac{u'}{M}\frac{v'}{M}\,.
\end{equation}
All of this is done so that  $u'$ and $v'$
have ranges $-\infty<u'<\infty$ and $-\infty<v'<\infty$,
which Eqs.~(\ref{metricu'v'}) and (\ref{rimplicitu'v'}) permit.
To obtain Eqs.~(\ref{metricu'v'}) and (\ref{rimplicitu'v'})
directly from the $E\to\infty$ limit of Eqs.~(\ref{eqn:metricE>1})
and~(\ref{rimplicitEoo}), respectively, see
the Appendix.
Several properties are worth mentioning.

\vskip 0.5cm
In terms of the coordinates $(u,v)$, or
$(t,r)$, the coordinate
transformations that yield 
the maximal extended null coordinates 
$(u',v')$
with infinite ranges have to be broadened, resulting
in the existence of four regions, regions I, II, III, and IV.
Region I is the region where 
the transformations 
Eqs.~(\ref{udashed}) and~(\ref{vdashed})
hold, i.e., it is a region with $u'\leq 0$ and
$v'\geq 0$, or a region with  $r\geq 2M$ and $-\infty<t<\infty$.
Region II, a region for which $r\leq 2M$, gets a different set of
coordinate transformations.  In this region $r\leq 2M$, due to the
moduli appearing in Eqs.~(\ref{uedd}) and~(\ref{vedd}) and the change
of sign in Eq.~(\ref{metricuv}), one defines instead $u'$ as
$\frac{u'}{M}=+ \exp{\left(-\frac{u}{4M}\right)}=
\sqrt{1-\frac{r}{2M}}\, \exp{\left(-\frac{t}{4M}+\frac{r}{4M}\right)}$
and $v'$ as $\frac{v'}{M}=\exp{\left(\frac{v}{4M}\right)}=
\sqrt{1-\frac{r}{2M}}\, \exp{\left(
\frac{t}{4M}+\frac{r}{4M}\right)}$. These transformations are valid
for the region with $u'\geq 0$ and $v'\geq 0$, or the region with
$r\leq2M$ and $-\infty<t<\infty$.  Note that the coordinate
transformations in this region give $\left(1-\frac{r}{2M}\right)
\exp{\left(\frac{r}{2M}\right)}=\frac{u'}{M}\frac{v'}{M}$.  But this
is automatically incorporated into Eq.~(\ref{rimplicitu'v'}), so there
is no further concern on that.
Region III is another $r\geq 2M$ region.  Now one defines
$\frac{u'}{M}= \exp{\left(-\frac{u}{4M}\right)}=
\sqrt{\frac{r}{2M}-1}\, \exp{\left(-\frac{t}{4M}+\frac{r}{4M}\right)}$
and $v'$ as $\frac{v'}{M}=-\exp{\left(\frac{v}{4M}\right)}=-
\sqrt{\frac{r}{2M}-1}\, \exp{\left(
\frac{t}{4M}+\frac{r}{4M}\right)}$.  These transformations are valid
for the region with $u'\geq 0$ and $v'\leq 0$, or the region with
$r\geq2M$ and $-\infty<t<\infty$.  Note that the coordinate
transformations in this region give $\left(\frac{r}{2M}-1\right)
\exp{\left(\frac{r}{2M}\right)}=-\frac{u'}{M}\frac{v'}{M}$.  But this
is automatically incorporated into Eq.~(\ref{rimplicitu'v'}), so again
there is no further concern on that.
Region IV is another region with $r\leq 2M$.  Now, one defines $u'$ as
$\frac{u'}{M}=- \exp{\left(-\frac{u}{4M}\right)}= \allowbreak =-
\sqrt{1-\frac{r}{2M}}\, \exp{\left(-\frac{t}{4M}+\frac{r}{4M}\right)}$
and $v'$ as $\frac{v'}{M}=-\exp{\left(\frac{v}{4M}\right)}=-
\sqrt{1-\frac{r}{2M}}\, \exp{\left(
\frac{t}{4M}+\frac{r}{4M}\right)}$.  These transformations are valid
for the region with $u'\leq 0$ and $v'\leq 0$, or the region with
$r\leq2M$ and $-\infty<t<\infty$.  The coordinate transformations in
this region give as well $\left(1-\frac{r}{2M}\right)
\exp{\left(\frac{r}{2M}\right)}=\frac{u'}{M}\frac{v'}{M}$.  But this
is automatically incorporated into Eq.~(\ref{rimplicitu'v'}), so once
again there is no further concern on that.

\vskip 0.5cm
Furthermore, from Eq.~(\ref{rimplicitu'v'}) we see that the event
horizon at $r=2M$ has two solutions, $u'=0$ and $v'=0$ which are null
surfaces represented by straight lines. The true curvature singularity
at $r=0$ has two solutions $\frac{u'}{M}\frac{v'}{M}=1$, i.e., two
spacelike hyperbolae.  Implicit in the construction, there is a
wormhole, or Einstein-Rosen bridge, topology, with its throat
expanding and contracting.  The dynamic wormhole is non traversable,
but it spatially connects region I to region III through regions II
and IV.  Regions I and III are two asympotically flat regions causally
separated, region II is the black hole region, and region IV is the
white hole region of the spacetime.

\begin{figure}[h]
\centering
\subfigure[]{
\includegraphics[scale=0.3]{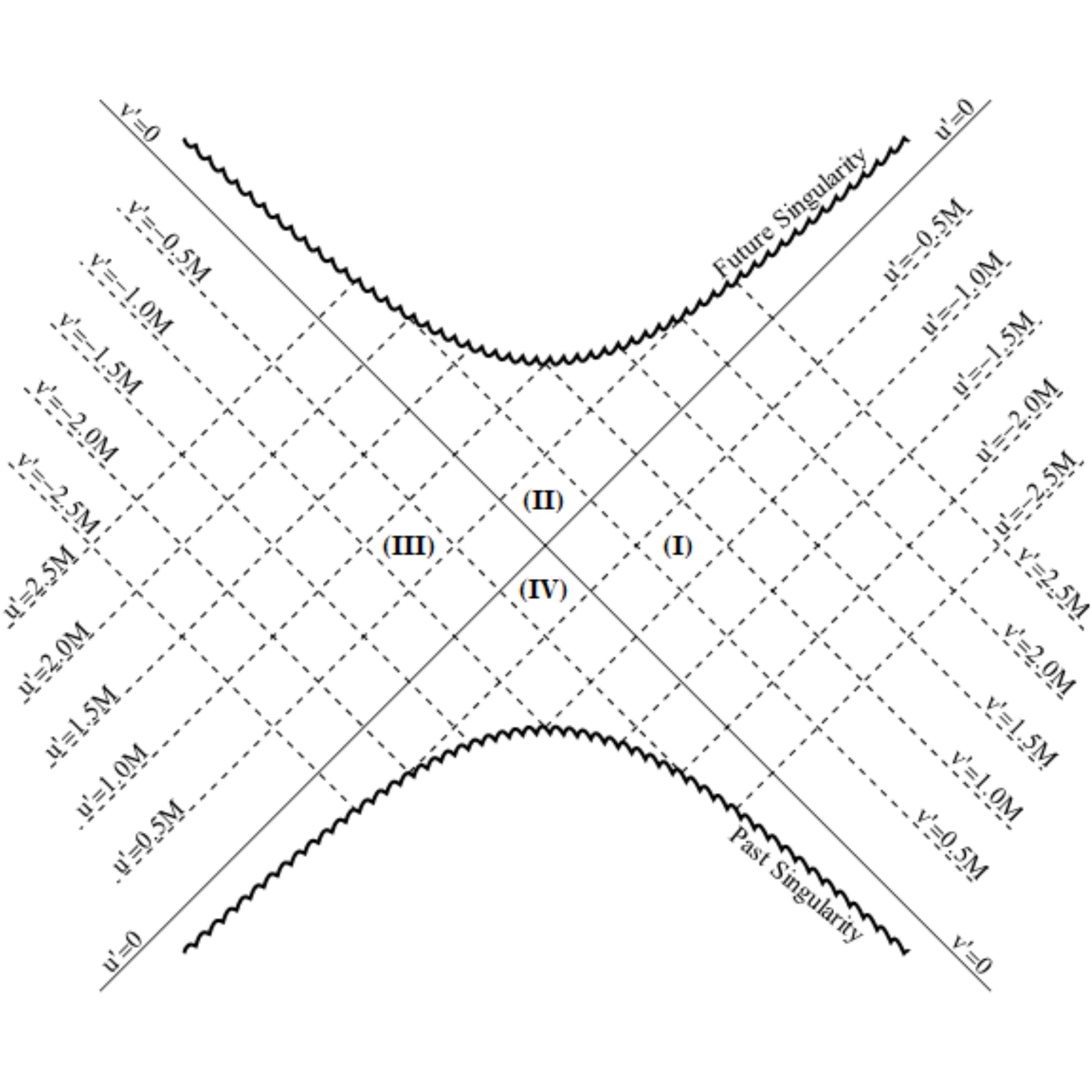}
\label{fig:diagram_tau_rhoa}
}\qquad
\subfigure[]{
\includegraphics[scale=0.3]{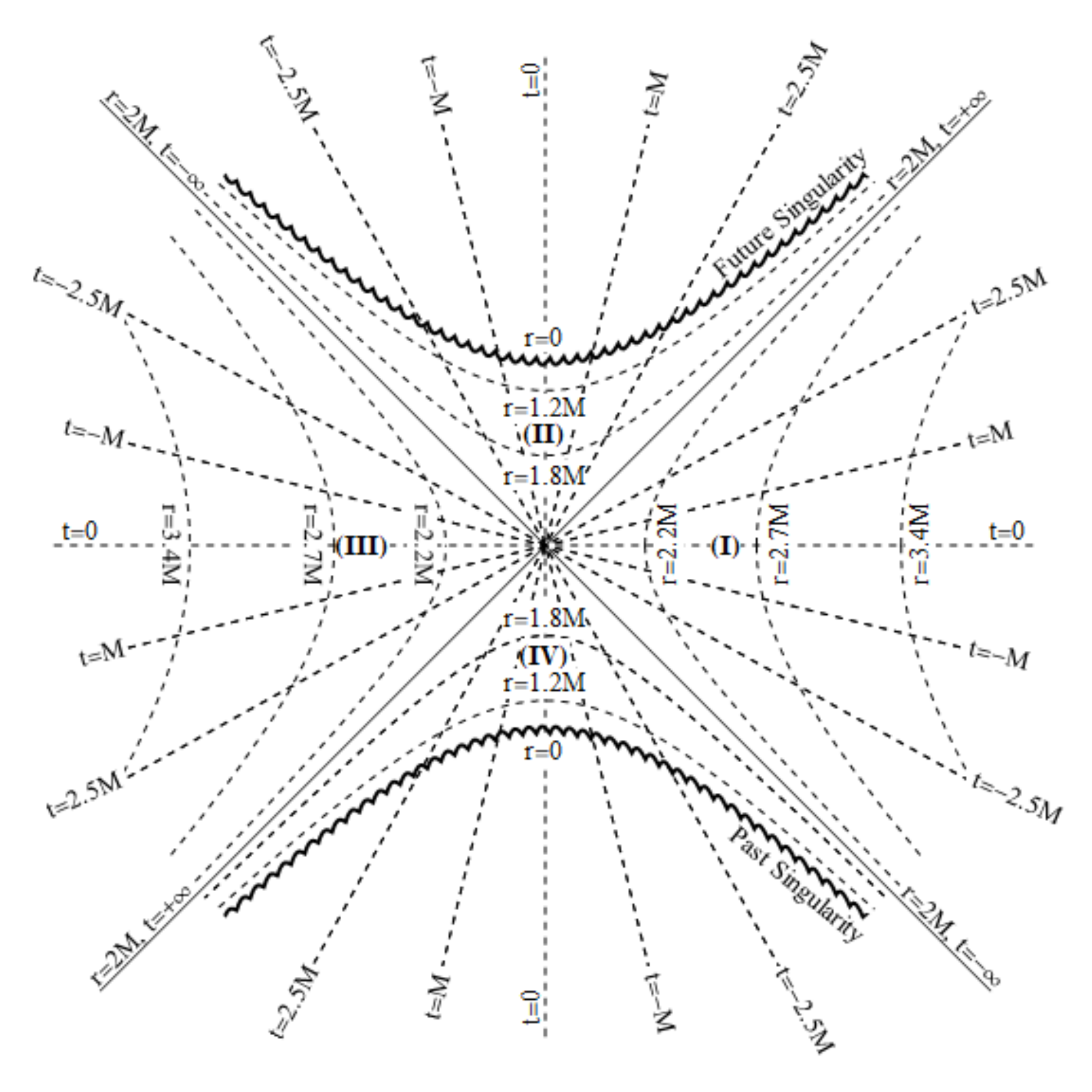}
\label{fig:diagram_tau_rhob}
}
\caption{
The maximal analytical extension of the Schwarzschild metric for the
parameter $E=\infty$, i.e., the Kruskal-Szekeres maximal extension, in
the plane $(u',v')$ is shown in a diagram with two different
descriptions, (a) and (b). In (a) typical values for lines of constant
$u'$ and constant $v'$ are displayed.  In (b) typical values for lines
of constant $t$ and constant $r$ are displayed.  The diagram, both
in (a) and in (b), represents a spacetime with
a wormhole, not shown, that forms out of
a singularity in the white hole region, i.e., region IV, and finishes
at the black hole region and its singularity, i.e., region II,
connecting the two separated asymptotically flat spacetimes, regions I
and III.  The $E=\infty$ diagram, i.e., the Kruskal-Szekeres diagram,
is very similar to the $E>1$ generic case diagram, see
Figure~\ref{fig:diagramEgt1}, as it is expected for a maximal
extension of the Schwarzschild spacetime, in particular for those
extensions within the same family.
}
\label{fig:diagram_tau_rho}
\end{figure}

\vskip 0.5cm
Eqs.~(\ref{metricu'v'}) and~(\ref{rimplicitu'v'}) together with the
corresponding interpretation give the maximal extension of the
Schwarzschild metric for $E=\infty$, taken as the limit of $E>1$, in
the coordinates $(u',v',\theta,\phi)$. Of course, this the
Kruskal-Szekeres maximal analytical extension, now seen as the
$E=\infty$ member of the family of extensions of $E>1$.  Recalling
that $u'={\mathcal t}'|_{E=\infty}$ and $v'=\tau'|_{E=\infty}$, we see
that the two timelike congruences that specify the two analytically
extended time coordinates $\mathcal t'$ and $\tau'$ that yield the
maximal extension for $E>1$ turned into the two analytically extended
null retarded and advanced congruences $u'$ and $v'$ of the
Kruskal-Szekeres maximal extension.
The two-dimensional part
$(u',v')$ of the coordinate system $(u',v',\theta,\phi)$
is shown in Figure~\ref{fig:diagram_tau_rho} , both
for lines of constant $u'$ and constant $v'$ in part (a)
of the figure, and for lines of constant $t$ and constant $r$ in part
(b) of the figure, conjointly with the labeling of regions I, II, III,
IV, needed to cover it.

\newpage

It is also worth discussing the normals to the $u'={\rm constant}$ and
$v'={\rm constant}$ hypersurfaces.  For that, we see that from
Eq.~(\ref{eqn:metricE>1}) in the limit $E\to\infty$, or directly from
Eq.~(\ref{metricu'v'}), one finds that the metric has covariant
components $g_{u'u'}=0$, $g_{v'v'}=0$, $g_{u'v'}=g_{v'u'}=
-\frac{16M}{r}\exp{\left( -{\frac{r}{2M}} \right)}$, $g_{\theta
\theta}=r^2$, $g_{\phi \phi}=r^2\sin^2{\theta}$.
The contravariant components of the metric can be calculated to be
$g^{u'u'}=0$, $g^{v'v'}=0$, $g^{u'v'}=g^{v'u'}= -\frac{r}{16M}
\exp{\left( \frac{r}{2M}\right)}$, $g^{\theta \theta}=\frac{1}{r^2}$,
$g^{\phi \phi}=\frac{1}{r^2\sin^2{\theta}}$.
The normals $n_a$ to the
$u'={\rm constant}$ and $v'={\rm constant}$ hypersurfaces are
${n^{u'}}_a=(1,0,0,0)$ and ${n^{v'}}_a=(0,1,0,0)$, respectively, where
the superscripts $u'$ and $v'$ in this context are not indices, they
simply label the respective normal. Their contravariant components are
$n^{{u'}\hskip0.005cm a}=(g^{u'u'}, g^{u'v'},0,0) =(0,-\frac{r}{16M}
\exp{\left( \frac{r}{2M}\right)},0,0)$ and $n^{{v'}\hskip0.005cm a}=
(g^{v'u'},g^{v'v'},0,0) = (-\frac{r}{16M} \exp{\left(
\frac{r}{2M}\right)},0,0,0)$.  The norms are then
${n^{u'}}_an^{{u'}\hskip0.005cm a}=0$ and
${n^{v'}}_an^{{v'}\hskip0.005cm a}=0$, respectively.  Thus, clearly,
the normals to the $u'={\rm constant}$ and $v'={\rm constant}$
hypersurfaces are null, and so $u'$ and $v'$ are null
coordinates, and the corresponding hypersurfaces are null
as well.



\section{Causal diagrams from $E=1$ to $E=\infty$}
\label{allEm}

\begin{figure}[h]
\centering
\includegraphics[scale=0.3]{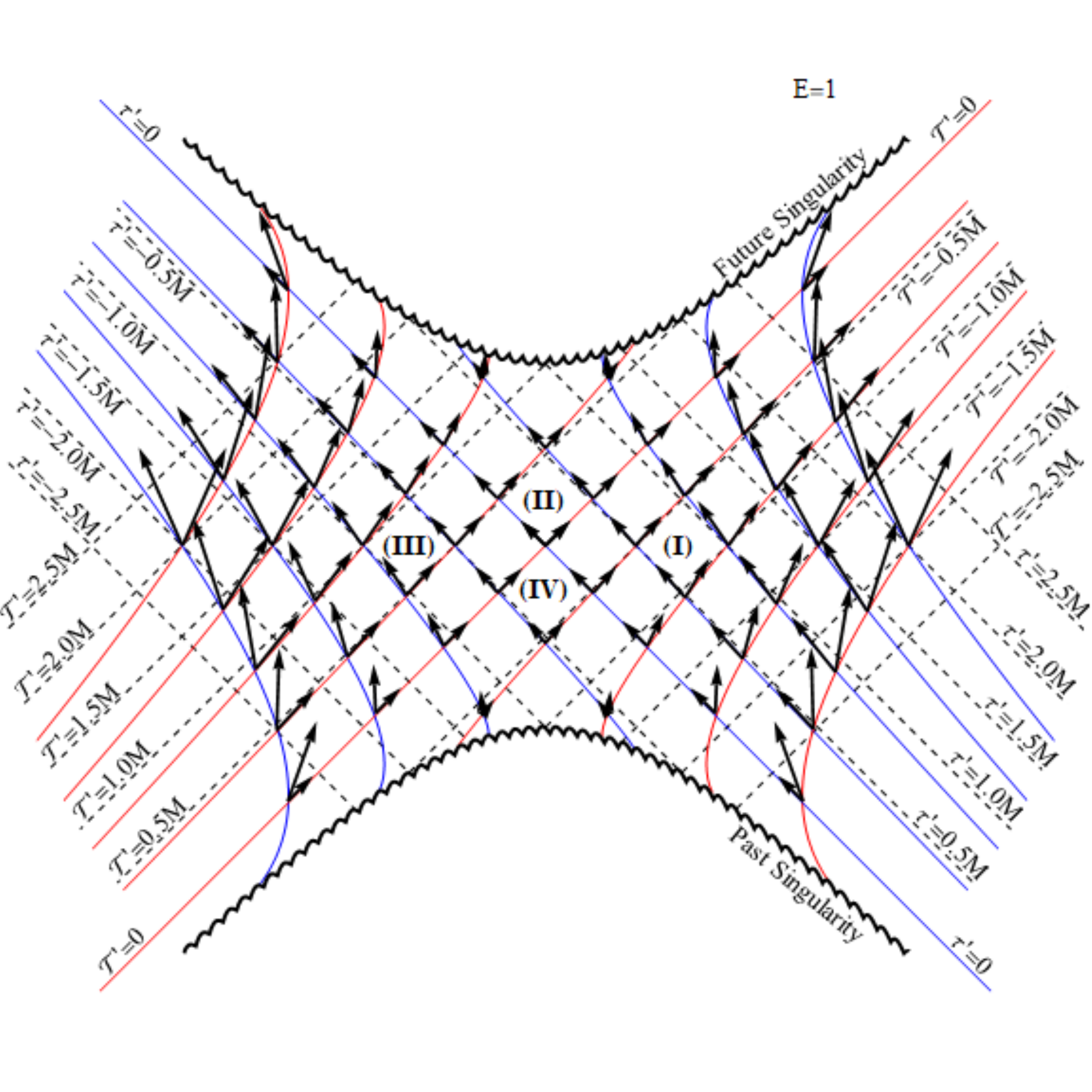}
\caption{Causal diagram for the maximal analytical extension in the
$E=1$ case. The two singularities and the two event horizons are shown
together with lines of outgoing and ingoing null geodesics, drawn in
red and blue, respectively, and with lines of constant ${\mathcal t}'$
and $\tau'$.  The contravariant normals to the ${\mathcal t}'={\rm
constant}$ and $\tau'={\rm constant}$
hypersurfaces, i.e., $n^{{\mathcal
t'}\hskip0.005cm a}$ and $n^{\tau'\hskip0.005cm a}$, respectively, are
also shown, with their timelike character clearly exhibited.
See text for more details.}
\label{fig:diagram_tau_rho_10}
\end{figure}

\begin{figure}[h]
\centering
\includegraphics[scale=0.3]{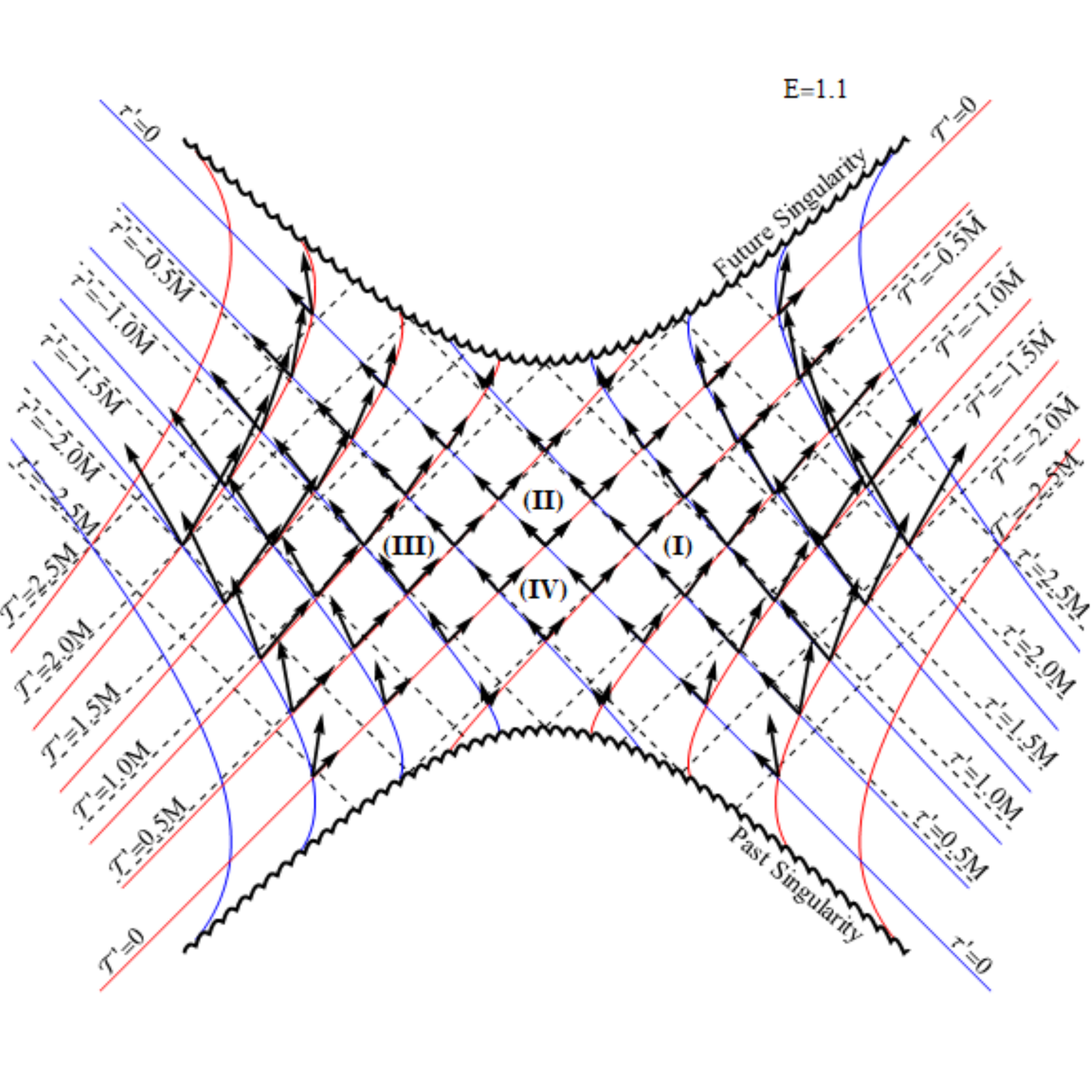}
\caption{Causal diagram for the maximal analytical extension in the
$E=1.1$ case. The two singularities and the two event horizons are
shown together with lines of outgoing and ingoing null geodesics,
drawn in red and blue, respectively, and with lines of constant
${\mathcal t}'$ and $\tau'$.  The contravariant normals to the
${\mathcal t}'={\rm constant}$ and $\tau'={\rm constant}$
hypersurfaces, i.e., $n^{{\mathcal t'}\hskip0.005cm a}$ and
$n^{\tau'\hskip0.005cm a}$, respectively, are also shown,
with their timelike character clearly exhibited.  See text
for more details.}
\label{fig:diagram_E11}
\end{figure}

\begin{figure}[h]
\centering
\includegraphics[scale=0.3]{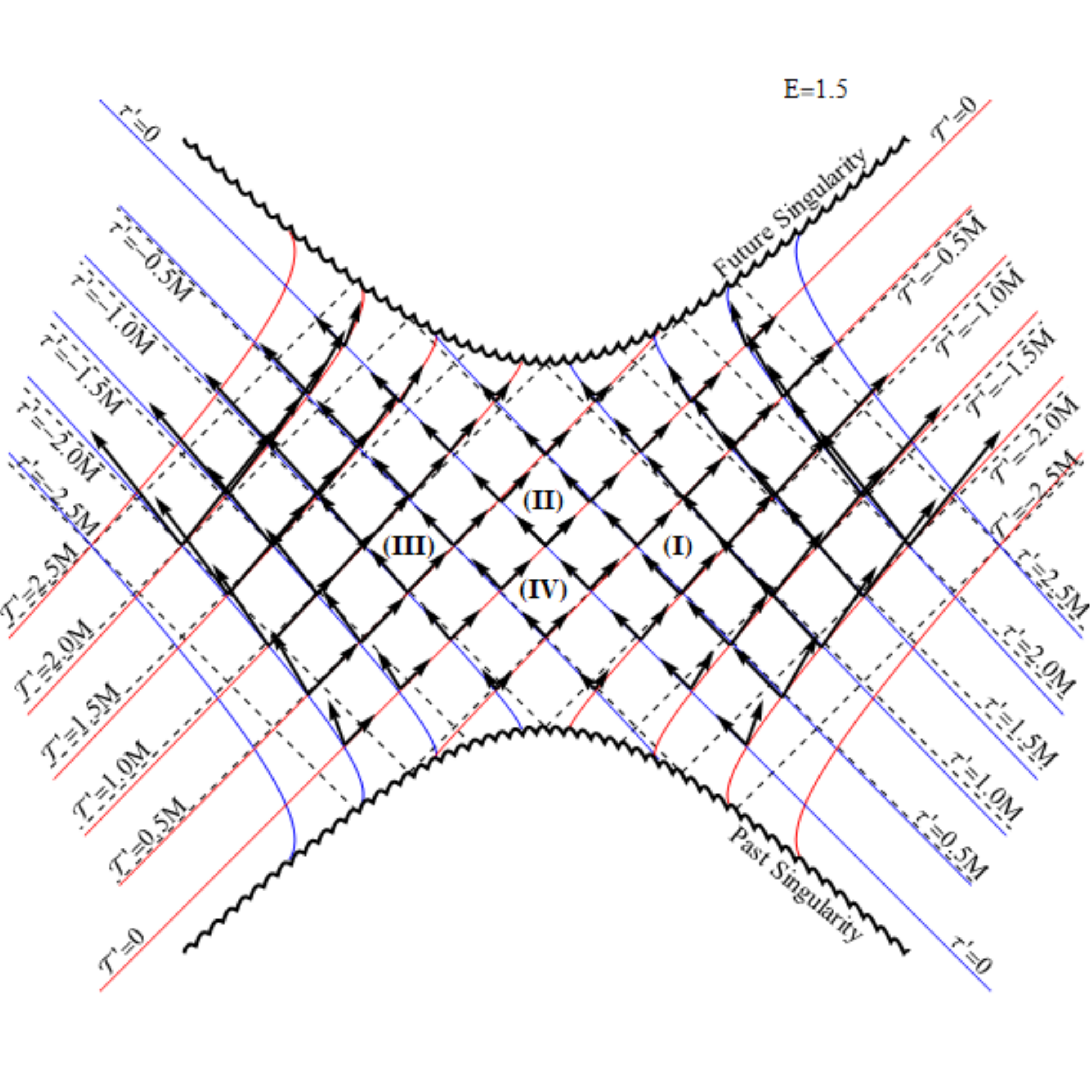}
\caption{Causal diagram for the maximal analytical extension in the
$E=1.5$ case. The two singularities and the two event horizons are
shown together with lines of outgoing and ingoing null geodesics,
drawn in red and blue, respectively, and with lines of constant
${\mathcal t}'$ and $\tau'$.  The contravariant normals to the
${\mathcal t}'={\rm constant}$ and $\tau'={\rm constant}$
hypersurfaces, i.e., $n^{{\mathcal t'}\hskip0.005cm a}$ and
$n^{\tau'\hskip0.005cm a}$, respectively, are also shown,
with their timelike character clearly exhibited.  See text
for more details.}
\label{fig:diagram_E15}
\end{figure}

\begin{figure}[h]
\centering
\includegraphics[scale=0.3]{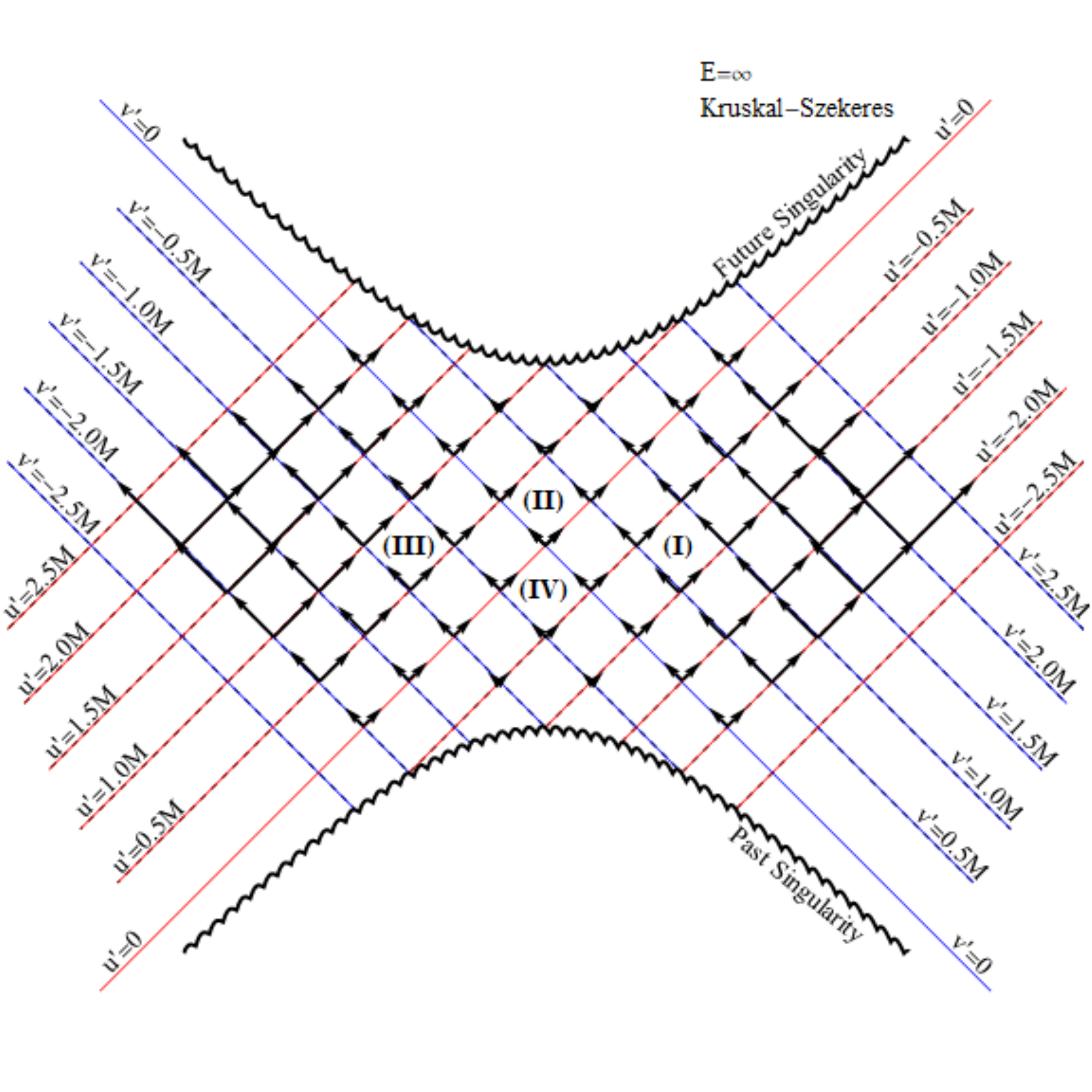}
\caption{Causal diagram for the maximal analytical extension in the
$E=\infty$ case, i.e., the Kruskal-Szekeres maximal extension.
The two singularities and the two event horizons are
shown together with lines of outgoing and ingoing null geodesics,
drawn in red and blue, respectively, and with lines of constant
$u'\equiv\lim_{E\to \infty} {\mathcal t'}$ and
$v'\equiv\lim_{E\to \infty} {\tau'}$. In this $E=\infty$ case
these two sets of lines coincide. 
The contravariant normals to the
$u'={\rm constant}$ and $v'={\rm constant}$
hypersurfaces, i.e., $n^{u'\hskip0.005cm a}$ and
$n^{v'\hskip0.005cm a}$, respectively, are also shown,
with their null character clearly exhibited.  See text
for more details.}
\label{fig:diagram_tau_rho_inf}
\end{figure}

In this unified account that carries maximal extensions
of the Schwarzschild metric along the parameter $E$,
it is of interest to
trace the radial null geodesics
for several values of the parameter $E$ itself, $1\leq E\leq\infty$,
in the plane characterized by the $({\mathcal t}',\tau')$
coordinates.
Null geodesics have $ds^2=0$ along them,
and if they are radial then also $d\theta=0$ and $d\phi=0$.
Using the line element given in Eq.~(\ref{eqn:metricE>1})
together with 
Eq.~(\ref{rimplicitEoo}), we can then trace the
radial null geodesics, and with it the
causal structure for each $E$, in
the corresponding maximally analytic extended
diagram. 
Figures~\ref{fig:diagram_tau_rho_10},~\ref{fig:diagram_E11},
\ref{fig:diagram_E15}, and~\ref{fig:diagram_tau_rho_inf}
are the maximal extended causal diagrams for 
$E=1.0$,
$E=1.1$,
$E=1.5$,
and
$E=\infty$,
respectively.
In the $E=1$ case one can take the
null geodesics directly
from Eqs.~(\ref{metricu'v'Eeq1})-(\ref{rimplicitu'v'Eeq1}),
and in the $E=\infty$ case, i.e., the Kruskal-Szekeres extension,
directly
from Eqs.~(\ref{metricu'v'})-(\ref{rimplicitu'v'}).

The features shown in the four figures are:
(i) The past and future spacelike singularities at $r=0$.
(ii) The regions I, II, III, and IV, described earlier.
(iii) The lines of ${\mathcal t}'={\rm constant}$ and $\tau'={\rm
constant}$, in the $E=\infty$ case these are the lines of $u'={\rm
constant}$ and $v'={\rm constant}$.
(iv) The outgoing null geodesics represented by red lines and the ingoing
null geodesics represented by blue lines.
(v) The contravariant normals to the ${\mathcal t}'={\rm constant}$
and $\tau'={\rm constant}$, i.e., $n^{{\mathcal t'}\hskip0.005cm a}$
and $n^{\tau'\hskip0.005cm a}$, respectively, as given in detail
previously.

\centerline{}
\vskip 4cm
\centerline{}

\newpage

As it had to be, the lines of ${\mathcal t}'={\rm constant}$ and
$\tau'={\rm constant}$ are tachyonic, i.e., spacelike hypersurfaces, a
feature clearly seen by comparison of these lines with the ingoing and
outgoing null geodesic lines, except for ${\mathcal t}'=0$ and $\tau'=0$
which are null lines representing the $r=2M$ event horizons of the
solution that separate regions I, II, III, and IV.  In the
$E=\infty$ case, i.e., Kruskal-Szekeres, the spacelike lines turn into
the null lines $u'={\rm constant}$ and $v'={\rm constant}$, with
$u'=0$ and $v'=0$ being the event horizons separating regions I,
II, III, and IV.  One also sees that the contravariant normals
$n^{{\mathcal t'}\hskip0.005cm a}$ and $n^{\tau'\hskip0.005cm a}$, are
always inside the local light cone, and so the coordinates ${\mathcal
t'}$ and $\tau'$ are timelike, except at the horizons where they are
null. In the $E=\infty$ case, i.e., Kruskal-Szekeres, the
contravariant normals $n^{u'\hskip0.005cm a}$ and $n^{v'\hskip0.005cm
a}$ are null vectors always, and so the coordinates $u'$ and $v'$
are null, i.e., the ${\mathcal
t'}$ and $\tau'$ timelike coordinates
turned into the $u'$ and $v'$ null coordinates.

\vfill

\section{Conclusions}
\label{conc}

The scenario for maximally extend the Schwarzschild metric is now
complete. Schwarzschild is the starting point.  In the usual
standard coordinates, also called
Schwarzschild coordinates, its extension past the sphere $r=2M$ is
cryptic, in any case is not maximal, and to exhibit it fully one needs
two coordinate patches, altogether making it very difficult to obtain
a complete interpretation.  Departing
from it, there is one branch alone,
namely, the Painlev\'e-Gullstrand branch that works either with
outgoing or with ingoing timelike congruences, or equivalently with
outgoing or ingoing test particles placed over them, parameterized by
their energy per unit mass $E$, and that in the $E\to\infty$ limit
ends in the Eddington-Finkelstein retarded or advanced null
coordinates, respectively. The Painlev\'e-Gullstrand branch, including
its Eddington-Finkelstein $E=\infty$ endpoint, partially extends the
Schwarzschild metric past $r=2M$, but it is not maximal, to have the
full solution one needs two coordinate patches, which again inhibits
the full interpretation of the solution.  Then, from
Painlev\'e-Gullstrand there are two bifucartion branches.  One branch
is the Novikov-Lema\^itre that uses the Painlev\'e-Gullstrand time
coordinate and an appropriate radial comoving coordinate.  This branch
extends the Schwarzschild metric past $r=2M$, is maximal in the
Novikov range $0<E<1$ and partial only in the Lema\^itre range
$1\leq E<\infty$, ending, in the $E\to\infty$ limit, in Minkowski.  The
other branch is the one we found here, with the two analytically
extended Painlev\'e-Gullstrand time coordinates, one related to
outgoing, the other to ingoing timelike congruences.  This branch
extends the Schwarzschild metric past $r=2M$, is maximal and valid for
$1\leq E <\infty$, and ends, for $E=\infty$, directly, or if
wished, via the two analytically extended Eddington-Finkelstein
retarded and advanced null coordinates, in the Kruskal-Szekeres
maximal extension. The maximally extended solutions of the
Schwarzschild metric allow for an easy and full interpretation of its
complex spacetime structure.

Indeed, whereas the partial extensions of the Schwarzschild metric are
of great interest to analyze gravitational collapse of matter and
physical phenomena involving black holes where a future event horizon
makes its appearance, and in certain instances to analyze time
reversal white hole phenomena, the maximal extensions deliver the full
solution, showing a model dynamic universe with two separate spacetime
sheets, containing a past spacelike singularity, with a white hole
region delimited by a past event horizon, that join at a dynamic
nontraversable Einstein-Rosen bridge, or wormhole whose throat expands
up to $r=2M$, to collapse into the inside of a future event horizon
containing a black hole region with a future spacelike singularity
separating again the two separate spacetime sheets of this model
universe.  Here, a family of maximal extensions of the Schwarzschild
spacetime parameterized by the energy per unit mass $E$ of congruences
of outgoing and ingoing timelike geodesics has been obtained.  In this
unified description, the Kruskal-Szekeres maximal extension of sixty
years ago is seen here as the important, but now particular, instance
of this $E$ family, namely, the one with $E=\infty$. This maximal
description provides the link between Gullstrand-Painlev\'e and
Kruskal-Szekeres.

\begin{acknowledgments}
We acknowledge FCT - Funda\c c\~ao para a Ci\^encia e Tecnologia
of Portugal for financial support through
Project No.~UIDB/00099/2020.
\end{acknowledgments}


\section*{Appendix: Details for the $E\to 1$ limit
and the $E\to \infty$ Kruskal-Szekeres limit
from the $E>1$ generic case}
\label{app}

In order to see the continuity of the maximal extension
parameterized by $E$, we take the
generic $E>1$ case, and from it
obtain directly the limit to the case
$E=1$, and  the limit to the case $E=\infty$, i.e., the
Kruskal-Szekeres extension.

\vskip 0.8cm

\noindent
{\it $E=1$ limit from $E>1$:}
\vskip 0.2cm
\noindent
Here we take the $E\to 1$ limit of 
Eqs.~(\ref{eqn:metricE>1}) and~(\ref{rimplicitEoo}).
We will do it term by term in each equation.
For Eq.~(\ref{eqn:metricE>1}) we have:
$\lim_{E\to 1} -4 \left( \frac{ 2E^2-1+ \frac{2M}{r}+ 2E
\sqrt{E^2-1+\frac{2M}{r}}}{ E^2-1+\frac{2M}{r}} \right)=
-4 \frac{(1+ \sqrt{\frac{2M}{r}})^2}{\frac{2M}{r}}$;
$\lim_{E\to 1} \exp{\left( -\frac{r}{2ME} \sqrt{E^2-1+
\frac{2M}{r}} \right)}=\exp{\left( -\sqrt{\frac{r}{2M}} \right)}$;
$\lim_{E\to 1} \left( \frac{\frac{M}{r}}{E^2-1+ \frac{M}{r}+
\sqrt{E^2-1}\sqrt{E^2-1+\frac{2M}{r}}} \right) ^{\frac{2E^2-1}{2E
\sqrt{E^2-1}}} =
\left(1+2\sqrt{E^2-1}
\sqrt{\frac{r}{2M}}\right)^{-\frac{1}{2\sqrt{E^2-1}}}
=\exp\left[-\frac{1}{2\sqrt{E^2-1}}
\ln\left(1+2\sqrt{E^2-1}
\right.\right.\times
\\
\left.\left.\times
\sqrt{\frac{r}{2M}}\right)
\right]
=
\exp{\left( -\sqrt{\frac{r}{2M}} \right)}$;
$\lim_{E\to 1}
-\frac{1}{M^2} \left(2E^2-1+ \frac{2M}{r}+2E
\sqrt{E^2-1+ \frac{2M}{r}} \right)
= -\frac{1}{M^2} \left( 1+ \sqrt{\frac{2M}{r}} \right)^2$;\\
$\lim_{E\to 1} \exp\left( -\frac{r}{2ME}
\sqrt{E^2-1+
\frac{2M}{r}} \right)=
\exp{\left( -\sqrt{\frac{r}{2M}} \right)}$;
$\lim_{E\to 1}
\left(
\frac{\frac{M}{r}}{E^2-1+\frac{M}{r}+
\sqrt{E^2-1} \sqrt{E^2-1+\frac{2M}{r}}}
\right)^{\frac{2E^2-1}{2E \sqrt{E^2-1}}}
=\exp{\left( -\sqrt{\frac{r}{2M}} \right)}$;
$\lim_{E\to 1}
2 \left( 2E^2-1+
\frac{2M}{r} \right)
=2 \left( 1+ \frac{2M}{r} \right)$.
Thus, Eq.~(\ref{eqn:metricE>1}) is now
$\mathrm{d}s^2=-4 \frac{(1+
\sqrt{\frac{2M}{r}})^2}{\frac{2M}{r}} \exp{\left( -2
\sqrt{\frac{r}{2M}} \right)} \Big[ -\frac{1}{M^2} \left( 1+
\sqrt{\frac{2M}{r}} \right)^2\times 
\\
\times \exp{\left( -2 \sqrt{\frac{r}{2M}}
\right)} (\tau'^2 \mathrm{d} {\mathcal t}'^2+ {\mathcal t}'^2
\mathrm{d} \tau'^2)+ 2\left(1+ \frac{2M}{r} \right) \mathrm{d}
{\mathcal t}'\, \mathrm{d} \tau'^2 \Big]+ r^2 (\mathrm{d} \theta^2+
\sin^2{\theta}\, \mathrm{d} \phi^2)$.
This is the line element found for the $E=1$ case, see
Eq.~(\ref{metricu'v'Eeq1}).
For Eq.~(\ref{rimplicitEoo}) we have:
$\lim_{E\to 1}\left( \frac{
\frac{r}{2M}-1}{2E^2-1+ \frac{2M}{r}+ 2E \sqrt{E^2-1+
\frac{2M}{r}}} \right) \frac{2M}{r}=
\frac{\sqrt{\frac{r}{2M}}- 1}{\sqrt{\frac{r}{2M}}+ 1}$;
$\lim_{E\to 1}
\exp{\left( \frac{r}{2ME}
\sqrt{E^2-1+ \frac{2M}{r}}  \right)}=
\exp{\left( \sqrt{\frac{r}{2M}} \right)}$;
$\lim_{E\to 1}
\left[
\frac{r}{M}
\left( E^2-1 + \frac{M}{r}+ \sqrt{E^2-1}
\sqrt{E^2-1+ \frac{2M}{r}}\right)
\right]^{\frac{2E^2-1}{2E \sqrt{E^2-1}}}
=\allowbreak
=\left(1+2\sqrt{E^2-1}
\sqrt{\frac{r}{2M}}\right)^{\frac{1}{2\sqrt{E^2-1}}}
=\exp\left[\frac{1}{2\sqrt{E^2-1}}
\ln\left(1+2\sqrt{E^2-1}
\sqrt{\frac{r}{2M}}\right)
\right]
=
\exp{\left( \sqrt{\frac{r}{2M}} \right)}$.
Thus, Eq~(\ref{rimplicitEoo}) in the
$E\to1$ limit turns into
$\frac{\sqrt{\frac{r}{2M}}-1}{\sqrt{\frac{r}{2M}}+1}
\exp{\left( 2 \sqrt{\frac{r}{2M}} \right)}=
-\frac{{\mathcal t}'}{M}
\frac{\tau'}{M}
$. 
This is indeed Eq.~(\ref{rimplicitu'v'Eeq1}).

\vskip 0.8cm

\noindent
{\it $E\to \infty$ limit from $E>1$, the Kruskal-Szekeres
line element:}
\vskip 0.2cm
\noindent
Here we take the $E\to \infty$ limit of 
Eqs.~(\ref{eqn:metricE>1}) and~(\ref{rimplicitEoo}).
We will do it term by term in each equation.
For Eq.~(\ref{eqn:metricE>1}) we have:
$\lim_{E\to \infty} -4 \left( \frac{ 2E^2-1+ \frac{2M}{r}+ 2E
\sqrt{E^2-1+\frac{2M}{r}}}{ E^2-1+\frac{2M}{r}} \right)=-16$;
$\lim_{E\to \infty} \exp{\left( -\frac{r}{2ME} \sqrt{E^2-1+
\frac{2M}{r}} \right)}=\exp{\left( -\frac{r}{2M}\right)}$;
$\lim_{E\to \infty} \left(\frac{M}{r}\, \frac{1}{E^2-1+ \frac{M}{r}+
\sqrt{E^2-1}\sqrt{E^2-1+\frac{2M}{r}}} \right) ^{\frac{2E^2-1}{2E
\sqrt{E^2-1}}} =\frac{M}{r}\frac{1}{2E^2}$;
$\lim_{E\to \infty}
-\frac{1}{M^2} \left(2E^2-1+ \frac{2M}{r}+2E
\sqrt{E^2-1+ \frac{2M}{r}} \right)
=\allowbreak =-\frac{4E^2}{M^2}$;
$\lim_{E\to \infty} \exp{\left( -\frac{r}{2ME} \sqrt{E^2-1+
\frac{2M}{r}} \right)}=\exp{\left( -\frac{r}{2M}\right)}$;
$\lim_{E\to \infty}
\left(
\frac{\frac{M}{r}}{E^2-1+\frac{M}{r}+
\sqrt{E^2-1} \sqrt{E^2-1+\frac{2M}{r}}}
\right)^{\frac{2E^2-1}{2E \sqrt{E^2-1}}}
=\allowbreak =\frac{M}{r}\frac{1}{2E^2}$;
$\lim_{E\to \infty}
2 \left( 2E^2-1+ \frac{2M}{r} \right)
=4E^2$.
Thus, Eq~(\ref{eqn:metricE>1}) is now
$\mathrm{d}s^2=
-16\exp{\left( -\frac{r}{2M}\right)}
\frac{M}{r}\frac{1}{2E^2}
\Big[
-\frac{4E^2}{M^2}\exp{\left( -\frac{r}{2M}\right)}
\frac{M}{r}\times \\
\times \frac{1}{2E^2}
(v'^2\mathrm{d}u'^2+u'^2
\mathrm{d}v'^2)+
4E^2
\mathrm{d}u'\mathrm{d}v'\,
\Big]+ r^2 (\mathrm{d}\theta^2+
\sin^2{\theta}\, \mathrm{d}\phi^2)$,
where for convenience of notation
we have redefined the
coordinates, $u'\equiv{\mathcal t}'$
and $v'\equiv\tau'$. Implementing
definitely the $E\to\infty$ limit, the
term in $(v'^2\mathrm{d}u'^2+u'^2
\mathrm{d}v'^2)$ vanishes and one gets,
$\mathrm{d}s^2=-\frac{32M}{r}\exp{\left( -\frac{r}{2M}\right)}
\mathrm{d}u'\mathrm{d}v'+ r^2 (\mathrm{d}\theta^2+
\sin^2{\theta}\, \mathrm{d}\phi^2)$. This is Eq.(\ref{metricu'v'}),
i.e., the Kruskal-Szekeres line element.
For Eq.~(\ref{rimplicitEoo}) we have:
$\lim_{E\to \infty}\left( \frac{
\frac{r}{2M}-1}{2E^2-1+ \frac{2M}{r}+ 2E \sqrt{E^2-1+
\frac{2M}{r}}} \right) \frac{2M}{r}=
\frac{\frac{r}{2M}-1}{4E^2}\frac{2M}{r}$;
$\lim_{E\to \infty}
\exp{\left( \frac{r}{2ME}
\sqrt{E^2-1+ \frac{2M}{r}}  \right)}= \allowbreak =
\exp{\left( \frac{r}{2M} \right)}$;
$\lim_{E\to \infty}
\left[
\frac{r}{M}
\left( E^2-1 + \frac{M}{r}+ \sqrt{E^2-1}
\sqrt{E^2-1+ \frac{2M}{r}}\right)
\right]^{\frac{2E^2-1}{2E \sqrt{E^2-1}}}
=\dfrac{r}{M}\,2E^2$.
Thus, redefining for convenience of notation
the
coordinates ${\mathcal t}'$ and $\tau'$ as $u'\equiv{\mathcal t}'$
and $v'\equiv\tau'$, Eq~(\ref{rimplicitEoo}) in the
$E\to\infty$ limit turns into
$\left(\frac{r}{2M}-1\right)
\exp{\left( \frac{r}{2M} \right)}=
-\frac{u'}{M}
\frac{v'}{M}
$. 
This is Eq.~(\ref{rimplicitu'v'}),
i.e., the Kruskal-Szekeres
implicit definition of $r$ in terms of
$u'$ and $v'$.
Seen through this direct limiting procedure,
the Kruskal-Szekeres solution is
indeed a particular case of the $E$ family
of maximal extensions. In no place there
was explicit need to resort to 
Eddington-Finkelstein null coordinates and
their analytical extended versions.


\end{document}